\documentclass[superscriptaddress,preprintnumbers,nofootinbib,11pt]{revtex4-2}
\usepackage{amsmath,amssymb}
\usepackage[dvipdf,dvips]{graphicx}
\usepackage{color}
\usepackage{hyperref}
\usepackage{url}
\usepackage{slashed}
\usepackage{subfigure}
\usepackage[usenames,dvipsnames]{xcolor}
\usepackage{amsmath}
\usepackage{amsfonts}
\usepackage{float} 
\usepackage{amssymb}
\usepackage{epsfig}
\usepackage{graphics}
\usepackage{euscript}
\usepackage{slashed}
\usepackage{epstopdf}
\usepackage[utf8]{inputenc}
\usepackage{palatino}
\usepackage{bbold}
\usepackage[normalem]{ulem}
\usepackage{MnSymbol}
\allowdisplaybreaks

\newcommand{\be}{\begin{equation}}
\newcommand{\ee}{\end{equation}}
\newcommand{\bea}{\begin{eqnarray}}
\newcommand{\eea}{\end{eqnarray}}
\newcommand{\bnabla}{\bar\nabla}

\newcommand{\bg}{\bar g}

\newcommand{\Tr}{\mathrm{Tr}}

\newcommand{\tth}{{h^{TT}}}

\newcommand{\bphi}{\bar\phi}
\newcommand{\bR}{\bar R}
\newcommand{\cV}{{\cal V}}

\usepackage{verbatim}
\usepackage{graphicx}
\usepackage{latexsym}

\newcommand{\diff}{\mathit{Diff}}
\newcommand{\sdiff}{\mathit{SDiff}}

\def \Tr{\textmd{Tr}}

\def \and{\textmd{and}}

\def \be{\begin{equation}}
\def \ee{\end{equation}}

\def \bea{\begin{eqnarray}}
\def \eea{\end{eqnarray}}

\begin{document}

\title{Can quantum fluctuations
differentiate between standard and unimodular gravity?} 

\author{Gustavo P. de Brito} \email{gustavo@cp3.sdu.dk}
\affiliation{CP3-Origins, University of Southern Denmark, Campusvej 55, DK-5230 Odense M, Denmark}

\author{Oleg Melichev} \email{omeliche@sissa.it}
\affiliation{International School for Advanced Studies, via Bonomea 265, I-34136 Trieste, Italy and
INFN, Sezione di Trieste, Italy}

\author{Roberto Percacci} \email{percacci@sissa.it}
\affiliation{International School for Advanced Studies, via Bonomea 265, I-34136 Trieste, Italy and
INFN, Sezione di Trieste, Italy}

\author{Antonio D. Pereira} \email{A.Duarte@science.ru.nl and adpjunior@id.uff.br}
\affiliation{Institute for Mathematics, Astrophysics and Particle Physics 
(IMAPP),
Radboud University, Heyendaalseweg 135, 6525 AJ Nijmegen, The Netherlands}
\affiliation{Instituto de F\'isica, Universidade Federal Fluminense, Campus da Praia Vermelha, Av. Litor\^anea s/n, 24210-346, Niter\'oi, RJ, Brazil}

\begin{abstract}
We formally prove the existence of a quantization procedure that makes the path integral of a general diffeomorphism-invariant theory of gravity, with fixed total spacetime volume, equivalent to that of its unimodular version. This is achieved by means of 
a partial gauge fixing of diffeomorphisms together with a careful definition of the unimodular measure. The statement holds also in the presence of matter. As an explicit example, we consider scalar-tensor theories and compute the corresponding logarithmic divergences in both settings. In spite of significant differences in the coupling of 
the scalar field to gravity, the results are equivalent for all couplings, including non-minimal ones. 
\end{abstract}

\maketitle

\section{Introduction}

General Relativity (GR) is in perfect agreement with all 
experimental data. Even if this is confirmed
by the next observational campaigns, 
another important issue remains open. 
Classical GR has several equivalent formulations
that may differ when quantum effects are taken into account. An interesting example is unimodular gravity (UG) 
\cite{Anderson:1971pn,vanderBij:1981ym,Buchmuller:1988wx,Buchmuller:1988yn,Weinberg:1988cp,Unruh:1988in,Unruh:1989db,Henneaux:1989zc,Ellis:2010uc,Ellis:2013uxa},
that we define as a theory of gravity constrained by
\be
\sqrt{|g|}=\omega\ ,
\label{unimodular}
\ee
where $\omega$ is a fixed volume form
\footnote{One often just takes $\omega=1$,
which justifies the name,
but this may not be applicable globally,
so we stick to the more general definition.}.
UG was advocated to be better suited to address some potential conceptual problems,
such as the cosmological constant problem and the problem of time in quantum gravity, offering, in this viewpoint, some advantages \cite{Wilczek:1983as,Brown:1989ne,Ng:1990rw,Ng:1990xz,Sorkin:1997gi,Alvarez:2007nn,Smolin:2009ti,Smolin:2010iq,Alvarez:2010cg,Carballo-Rubio:2015kaa,Alvarez:2015pla,Alvarez:2015sba,Percacci:2017fsy}, see, however, \cite{Fiol:2008vk,Padilla:2014yea,Padilla:2015aaa}.
At the classical level, GR and UG are ``almost equivalent'',
in the sense that (\ref{unimodular}) can be seen merely
as a gauge condition imposed on GR.
There is however a subtlety that prevents a full equivalence:
the identity
\be
\int \mathrm{d}^4x\sqrt{|g|}=\int \mathrm{d}^4x\,\omega\ ,
\label{unimodularint}
\ee
which follows from (\ref{unimodular}),
is a relation between observables,
and cannot be seen as a gauge condition\footnote{When the volume is infinite, 
one has to regulate it by ``putting the system in a box''
and impose (\ref{unimodularint}) on the regulated system.}. One must conclude that GR has one degree of freedom 
more than UG - just one, not one per spacetime point
\cite{Henneaux:1989zc}.
Thus, classical UG is equivalent to a version of classical GR 
where the total volume of spacetime is held fixed.
This can be seen in the Lagrangian formalism by adding to
the action a term
\be
\frac{\Lambda}{8\pi G}\left(V-\int \mathrm{d}^4x\sqrt{g}\right)
\label{volconstr}
\ee
where $\Lambda$ has to be thought of as a 
Lagrange multiplier enforcing that the
spacetime volume is equal to $V$\footnote{This point of view has been used by Hawking
in Euclidean quantum gravity, where he interpreted 
the resulting partition function as the ``volume canonical ensemble'', see \cite{Hawking:1979zw}}.
It is natural to expect that this distinction will affect the infrared
properties of the theory, but not its behavior at short scales.
In this paper we shall not discuss the global,
large scale properties and when we say ``GR'' we shall
implicitly mean ``GR with fixed total volume''.

It is then natural to ask whether this ``almost equivalence''
holds also for the quantum versions of the theories.
The unimodularity condition (\ref{unimodular}) 
restricts the invariance group
from diffeomorphisms ($\diff$) to special 
(volume-preserving) diffeomorphisms ($\sdiff$)
\footnote{In the recent literature, the group $\sdiff$
is often referred to as $\mathit{TDiff}$,
where $T$ stands for ``transverse''.},
and one may expect that this could lead to different quantum theories. 
For example, if we choose a unimodular gauge for GR, 
this requires a Faddeev-Popov determinant, 
while in the quantization of UG, the unimodularity condition 
is present \textit{ab initio} 
and no Faddeev-Popov determinants are necessary. 

In recent years, there appeared in the literature 
conflicting statements about the equivalence,
or lack thereof, between GR and UG at the quantum level, see, e.g., \cite{Smolin:2009ti,Smolin:2010iq,Alvarez:2015pla,Alvarez:2015sba,Bufalo:2015wda,Upadhyay:2015fna,Fiol:2008vk,Eichhorn:2013xr,Eichhorn:2015bna,Saltas:2014cta,Benedetti:2015zsw,Padilla:2014yea,Burger:2015kie,Alvarez:2016uog,Ardon:2017atk,Percacci:2017fsy,Gonzalez-Martin:2017fwz,Gonzalez-Martin:2018dmy,deBrito:2019umw,Yamashita:2020ixd,Baulieu:2020obv,Baulieu:2020rpv,Herrero-Valea:2020xaq,deBrito:2020rwu,deBrito:2020xhy}. 
We believe that some of these contradictions may be just
due to different quantization procedures.
In this work, we prove in general, based on formal path integral arguments,
that there exists a quantization procedure that preserves
the ``almost equivalence'' between these theories.
The proof goes through for any $\mathit{Diff}$-invariant action
and in this sense extends beyond ordinary GR.
Of course, there may be other definitions 
of the quantum theories that break the equivalence, 
but in the absence of other independent arguments 
in their favor, we think that the one we describe here is more natural.
Our argument is in the same spirit as the one presented in \cite{Fiol:2008vk,Padilla:2014yea} and extends the results of \cite{Ardon:2017atk,Percacci:2017fsy,Ohta:2018sze} beyond one-loop order. 
We should remark that both GR and UG are not renormalizable in perturbation theory and the formal 
path integrals should be ultraviolet (UV) regularized. Our formal proof relies on the use of the background field method, but we leave the parameterization of the metric, i.e., the way that we split the full metric in background and fluctuating parts, generic. Hence, this also extends previous results \cite{Ardon:2017atk,Percacci:2017fsy} which made explicit use of the so-called exponential split of the metric to impose the 
unimodularity condition \cite{Eichhorn:2013xr}. 

Our proof of equivalence is given initially for pure gravity
and one may again  worry that as soon as matter degrees of freedom are introduced, the equivalence would fall apart.
This is due to the different vertex structures.
In GR, the determinant of the metric produces infinitely many 
vertices between gravitons and matter fields that are absent in UG. 
Hence, Feynman rules are different in the two settings and one might expect that it is very unlikely that in the computation of an observable, miraculous cancellations lead to equivalent results. 
Yet, there are results in the literature explicitly showing
that this happens, 
see, e.g., \cite{Gonzalez-Martin:2017bvw,Gonzalez-Martin:2017fwz,Gonzalez-Martin:2018dmy}. 
In fact, we shall see that our formal proof of equivalence
extends also to the case when matter is present.

As an explicit check, we shall consider gravity non-minimally coupled to a scalar field and show that the one loop UV divergences are
the same for GR and UG.
This disagrees with \cite{Herrero-Valea:2020xaq},
who claimed that a particular dimensionless combination of couplings,
called $\Delta$, has different beta functions in the two settings.
In our calculation, the beta functions turn out to be the same.
What is perhaps more important, we find that the beta functions
of $\Delta$ are gauge-dependent, which may at least in part
explain the discrepancy.
Furthermore, the implementation of the unimodularity condition 
adopted in \cite{Herrero-Valea:2020xaq} is different from 
the one we use in this paper. 
We therefore think that the question whether different formulations 
of quantum UG can lead to different physical predictions than GR
remains still open.

The paper is structured as follows: In Sect.~\ref{Sect:EPI}, we define the path integral of diffeomorphism-invariant theories and formally show that it is possible to partially fix the gauge so as to reduce it to the one of UG. 
In Sect.~\ref{Sect:nmsct}, we perform an explicit computation in order to verify the claim of Sect.~\ref{Sect:EPI}
in the presence of matter. 
We consider scalar-tensor theories with a non-minimal coupling. 
We then proceed to the calculation of the one-loop beta functions in these theories both in the full diffeomorphism and special diffeomorphism invariant cases. 
In particular, we compute the running of $\Delta$ and discuss how it depends on the choice of gauge.
We also display the results of the running 
of $\Delta$ in GR with linear parameterization of the metric, 
in a generic linear covariant gauge, and compare our findings with the available literature. 
We collect our conclusions and perspectives in Sect~\ref{Concl}.
Appendix~\ref{betaFRGap} is mainly intended for users
of the functional renormalization group and explains
how to extract the logarithmic terms of the beta functions.
Appendix~ \ref{appB} contains some long expressions that are omitted in the main text. 

\goodbreak

\section{Equivalence of path integrals}\label{Sect:EPI}

The starting point of our analysis is the (Euclidean)\footnote{The Euclidean signature is not essential at this stage and the results could be equally deduced in the Lorentzian case.} path integral defined by a gravitational action $S_{\mathrm{Diff}}(g_{\mu\nu})$, 
$g_{\mu\nu} = g_{\mu\nu}(\bar{g};h)$ being the metric,
 $\bar{g}_{\mu\nu}$ a fixed background metric and $h_{\mu\nu}$ the fluctuating field which is integrated over. 
 The fluctuating field $h_{\mu\nu}$ does not need to be small, i.e., a perturbation around $\bar{g}_{\mu\nu}$. Moreover, the split of the full metric $g_{\mu\nu}$ in background and fluctuating parts is also general, not 
being restricted to the standard additive (linear) split. The action is assumed to be invariant under diffeomorphisms (but it is not restricted to 
be the Einstein-Hilbert action),
and so is the functional measure $\mathcal{D}h_{\mu\nu}$. Formally, the path integral is expressed as
\begin{equation}
\mathcal{Z}_{\mathrm{Diff}} = \int\frac{\mathcal{D}h_{\mu\nu}}{V_{\mathrm{Diff}}}\mathrm{e}^{-S_{\mathrm{Diff}}[g(\bar{g};h)]}\,.
\label{pi1}
\end{equation}
The factor $V_{\mathrm{Diff}}$ stands for the volume of the diffeomorphism group.

In most practical calculations within a continuum quantum-field theoretic 
setting, a gauge-fixing term must be introduced in \eqref{pi1}. This is typically achieved by the Faddeev-Popov procedure. The redundancy is generated by vector fields $\epsilon^\mu$ which can be decomposed as
\begin{equation}
\epsilon^{\mu} = \epsilon^\mu_{\mathrm{T}} + \nabla^{\mu}\phi\,,
\label{pi3}
\end{equation}
with $\nabla_\mu \epsilon^{\mu}_{\mathrm{T}}=0$ and $\nabla_\mu$  the covariant derivative defined with respect to the metric $g_{\mu\nu}$. The transverse vectorfields $\epsilon^{\mu}_{\mathrm{T}}$ generate the group \textit{SDiff}
of special (volume-preserving) diffeomorphisms.

Instead of introducing a single gauge-fixing condition for the entire group of diffeomorphism, 
we introduce two different conditions, first breaking \textit{Diff} to \textit{SDiff}, and then breaking \textit{SDiff}. 
This strategy has been discussed and worked out in a different way in \cite{Fiol:2008vk}, see also \cite{Padilla:2014yea}
and \cite{Ferrari:2013aza} for a general discussion of partial gauge fixing. 
In the first step we choose a gauge-fixing functional $\mathcal{F}(g)$ and insert the standard Faddeev-Popov identity given by
\begin{equation}
1=\Delta_\mathcal{F}(g)\int\mathcal{D}\phi~\delta(\mathcal{F}(g^{\phi}))\,,
\label{pi4}
\end{equation}
where $\Delta_\mathcal{F}(g)$ denotes the Faddeev-Popov determinant. 
The notation $g^{\phi}$ denotes the transformation of the metric generated by the longitudinal vectorfield
$\nabla_\mu\phi$:
\begin{equation}
\delta_\phi g_{\mu\nu} = 2\nabla_{\mu}\nabla_{\nu}\phi\,.
\label{pi5}
\end{equation}
We can now plug \eqref{pi4} in \eqref{pi1} leading to
\begin{equation}
\mathcal{Z}_{\mathrm{Diff}}=\int\frac{\mathcal{D}h_{\mu\nu}}{V_{\mathrm{Diff}}}\left(\Delta_\mathcal{F}(g)\int\mathcal{D}\phi~\delta(\mathcal{F}(g^{\phi}))\right)\mathrm{e}^{-S_{\mathrm{Diff}}[g(\bar{g};h)]}\,.
\label{pi6}
\end{equation}
Following the standard steps we now use the gauge invariance of the measure, of the Faddeev-Popov determinant and the action and redefine the integration variable, to get
\begin{equation}
\mathcal{Z}_{\mathrm{Diff}}=\int\frac{\mathcal{D}\phi\mathcal{D}h_{\mu\nu}}{V_{\mathrm{Diff}}}\Delta_\mathcal{F}(g)\delta(\mathcal{F}(g))\,\mathrm{e}^{-S_{\mathrm{Diff}}[g (\bar{g};h)]}\,.
\label{pi8}
\end{equation}
In \cite{Ardon:2017atk,Percacci:2017fsy} it was shown that 
\begin{equation}
V_{\mathrm{Diff}} = \mathrm{Det}\left(-\nabla^2\right)\times V_{\mathrm{SDiff}} \times \int\mathcal{D}\phi\,,
\label{pi9}
\end{equation}
where $V_{\mathrm{SDiff}}$ denotes the volume of the \textit{SDiff} group. Hence,
\begin{equation}
\mathcal{Z}_{\mathrm{Diff}} = \int\frac{\mathcal{D}h_{\mu\nu}}{V_{\mathrm{SDiff}}} \frac{1}{\mathrm{Det}\left(-\nabla^2\right)}\Delta_\mathcal{F}(g)\delta(\mathcal{F}(g))\mathrm{e}^{-S_{\mathrm{Diff}}[g (\bar{g};h)]}\,.
\label{pi10}
\end{equation}

An explicit example of this first stage of gauge fixing is the unimodular gauge defined by
\begin{equation}
\mathcal{F}(g) = \mathrm{det} g_{\mu\nu}-\omega^2(x)\,,
\label{pi11}
\end{equation}
$\omega(x)$ being a fixed scalar density. The delta function in \eqref{pi10} enforces that the full dynamical metric is unimodular. 
The corresponding Fadeev-Popov determinant is
\begin{equation}
\Delta_{\mathcal{F}}(g) = \mathrm{Det}\left(\omega^2(x)(-\nabla^2)\right)\,.
\label{pi12}
\end{equation} 
The contribution due to $\omega^2(x)$ in \eqref{pi12} can be absorbed in a normalization factor of the path integral and thereby it is harmless. Finally, by plugging \eqref{pi12} into \eqref{pi10}, yields
\begin{equation}
\mathcal{Z}_{\mathrm{Diff}} = \int\frac{\mathcal{D}h_{\mu\nu}}{V_{\mathrm{SDiff}}}\delta ( \mathrm{det} g_{\mu\nu}-\omega^2(x))\mathrm{e}^{-S_{\mathrm{Diff}}[g (\bar{g};h)]}\,.
\label{pi13}
\end{equation}
Due to the presence of the delta functional in \eqref{pi13}, the action in the Boltzmann factor collapses to its unimodular counterpart, i.e., $S_{\mathrm{Diff}}[g(\bar{g};h)]\to S_{\mathrm{SDiff}}[g(\bar{g};h)]$ where factors of $\sqrt{g}$ are replaced by $\omega(x)$ and when expanded in $h_{\mu\nu}$, the constraint $\mathcal{F}(g) = 0$ must be imposed. 
Eq.\eqref{pi13} is the path integral of UG with the unimodular measure $(\mathcal{D}h_{\mu\nu})_{\mathrm{UG}}$ defined by
\begin{equation}
(\mathcal{D}h_{\mu\nu})_{\mathrm{UG}} \equiv \mathcal{D}h_{\mu\nu}\,\delta ( \mathrm{det} g_{\mu\nu}-\omega^2(x))\,,
\label{pi14}
\end{equation}
i.e,
\begin{equation}
\mathcal{Z}_{\mathrm{Diff}} = \int\frac{(\mathcal{D}h_{\mu\nu})_{\mathrm{UG}}}{V_{\mathrm{SDiff}}}\mathrm{e}^{-S_{\mathrm{SDiff}}[g (\bar{g};h)]}\equiv\mathcal{Z}_{\mathrm{SDiff}}\,.
\label{pi15}
\end{equation}
One particular parameterization which is well-suited for the implementation of the unimodularity condition is the exponential split
\begin{equation}
g_{\mu\nu} = \bar{g}_{\mu\kappa}{(\mathrm{e}^{h})^{\kappa}}_{\nu}\,.
\label{pi16}
\end{equation}
Unimodularity of $g_{\mu\nu}$ is achieved by requiring the background to be unimodular 
($\mathrm{det}\bar{g} = \omega^2(x)$) 
and that the flucuations $h_{\mu\nu}$ are traceless\footnote{Another efficient method is the ``densitized" parameterization, see, e.g., \cite{Alvarez:2015sba,Ohta:2016npm}. If one opts for less efficient implementations, the unimodularity condition becomes difficult to implement in practical calculations. Nevertheless, for the partial gauge-fixing associated 
with the gauge freedom (\ref{pi5}), there seem to be no generation of quartic ghost terms \cite{Ferrari:2013aza} due to the fact that one just introduces a ghost-antighost pair.}.

In order to complete the gauge-fixing procedure, one applies again the Faddeev-Popov method for a gauge condition which fixes the $\textit{SDiff}$ 
invariance. This is achieved, e.g., by taking the standard linear covariant gauges in quantum gravity and applying the transverse projector to it. 
We refer to \cite{Alvarez:2008zw,Ardon:2017atk,Percacci:2017fsy,Eichhorn:2013xr,Eichhorn:2015bna,Benedetti:2015zsw,deBrito:2020rwu,deBrito:2020xhy} for more details. 

We remark that eq.\eqref{pi15} does not rely on the specific form of the gravitational action. Morever, if matter interactions were included (also 
of arbitrary form), the equivalence would still hold. In this case, the matter action $S^{\mathrm{Diff}}_{\mathrm{M}}(\varphi,\psi,A)$ is mapped to $S^{\mathrm{SDiff}}_{\mathrm{M}}(\varphi,\psi,A)$ with the replacement $\sqrt{g}\to\omega$ and fluctuations satisfying the constraint defined by 
the delta functional in \eqref{pi14}. Thus, we expect that gravity-matter 
systems in a full diffeomorphism-invariant setting are equivalent, quantum-mechanically, to gravity-matter systems in the unimodular framework. 

\section{Non-minimal comparisons in Scalar-tensor theories} \label{Sect:nmsct}

This section is devoted to the explicit calculation
of one-loop divergences in gravity-matter systems, illustrating
the quantum equivalence between \textit{Diff}-- and \textit{SDiff}-- invariant theories. In particular, we focus on scalar-tensor theories including non-minimal couplings between gravity and the scalar field. 

This system has been discussed 
recently in \cite{Herrero-Valea:2020xaq},
where it was claimed that through consideration of a suitable
dimensionless combination of couplings $\Delta$, it is possible to distinguish GR from UG. 
This would seem to contradict our results.
While for us the beta function $\Delta$ is the same in the
two theories, it turns out to be gauge dependent, 
thus weakening the significance of this test. 
We should also stress that in \cite{Herrero-Valea:2020xaq}, the authors employ a different way of implementing the unimodularity condition, 
and we do not exclude the existence of quantization schemes
that break the equivalence between GR and UG.

\subsection{Action}

The beta functions of GR coupled to a scalar have been derived previously 
in, e.g., \cite{Narain:2009fy} for the general class of actions
\begin{equation}
\label{action}
S [\phi,g] = \int\,\mathrm{d}^{d} x\sqrt{g}\,\left(V(\phi)-F(\phi)R+\frac{1}{2} \nabla_{\mu} \phi \nabla^\mu \phi \right)\,.
\end{equation}
This includes an arbitrary potential $V$ and arbitrary non-minimal couplings parametrized by the function $F$. If one expands $V$ and $F$ in Taylor series in $\phi$, with the additional assumption of invariance under $\phi\to -\phi$,
\bea
V(\phi)&=&\cV+\frac{1}{2}m^2\phi^2+\lambda\phi^4\ldots\ ,\quad
\cV=\frac{\Lambda}{8\pi G_N}
\label{vpol}
\\
F(\phi)&=&Z_N+\frac{1}{2}\xi\phi^2+\ldots\ ,\quad Z_N=\frac{1}{16\pi G_N}
\label{fpol}
\eea
We are especially interested in the dimensionless couplings $\xi$ and $\lambda$, whose leading one-loop beta functions are universal, and in dimensionless ratios of the dimensionful couplings, such as $G_N\Lambda$, $G_N 
m^2$, $\Lambda/m^2$, since their beta functions are also known to be less 
gauge- and parameterization-dependent. 
In \cite{Narain:2009fy}, the beta functions were computed by the use of the functional renormalization group (FRG) equation which is based on a cutoff-like regularization. Thus, power-law divergences are also taken into 
account. In \cite{Herrero-Valea:2020xaq},
on the other hand, the authors employed dimensional regularization which is blind to the power-law divergences.
For a direct comparison, we would have to extract from the FRG the ``universal" contributions, i.e., those related to logarithmic running. 
This is discussed in Appendix~\ref{uniFRG}.
In the next section we directly extract the beta functions from
the logarithmic divergences, calculated with heat kernel methods.

\subsection{Dynamical gravitons: UG or GR in exponential parametrization}

We start from GR
in the exponential parametrization \eqref{pi16} and 
follow the procedure of \cite{Percacci:2015wwa}.
We decompose the metric fluctuation in its irreducible spin 2, 1 and 0 components:
\begin{equation}
h_{\mu\nu} = h^{\mathrm{TT}}_{\mu\nu}+\bar{\nabla}_\mu\xi_\nu+\bar{\nabla}_\nu\xi_\mu+\bar{\nabla}_\mu \bar{\nabla}_\mu \sigma - \frac{1}{4}\bar{g}_{\mu\nu}\bar{\nabla}^{2}\sigma + \frac{1}{4}\bar{g}_{\mu\nu}h\,,
\label{yorkdecomp1}
\end{equation}
with $\bar{\nabla}^{\mu} h^{\mathrm{TT}}_{\mu\nu}=0$, $\bar{\nabla}^{\mu}\xi_\mu =0$ and $\bar{g}^{\mu\nu}h_{\mu\nu} = h$. 
A redefinition of the fields $\sigma$ and $\xi_\mu$ is performed in order 
to cancel the Jacobian generated by the York decomposition \eqref{yorkdecomp1},
\begin{equation}
\xi^\prime_\mu = \sqrt{-\bar{\nabla}^2-\frac{\bar{R}}{4}}\xi_\mu\,,\qquad\mathrm{and}\qquad\sigma^\prime = \sqrt{-\bar{\nabla}^2}\sqrt{-\bar{\nabla}^2-\frac{\bar{R}}{3}}\sigma\,.
\label{yorkdecomp2}
\end{equation}
We take the background metric $\bar{g}_{\mu\nu}$ to be a four-dimensional 
Euclidean maximally symmetric space. 
Then we choose the ``unimodular physical gauge'', which consists of setting to zero the spin one field $\xi'_\mu$ and the spin-0 field $h$.
With these choices, the gauge fixed Hessian is 
\bea
\label{gfhess0}
\tilde{S}^{(2)}_{\mathrm{grav}}&=&\int \mathrm{d}^4x\sqrt{\bg}\,\Biggl[
\frac{1}{4}F(\bphi)\tth_{\mu\nu}\left(-\bnabla^2+\frac{\bR}{6}\right)\tth^{\mu\nu}
-\frac{3}{32}F(\bphi)
\sigma'(-\bnabla^2)\sigma'
\nonumber\\
&-&\frac{3}{4}F'(\bphi)\delta\phi
\sqrt{(-\bnabla^2)\left(-\bnabla^2-\frac{\bR}{3}\right)}\sigma' +\frac{1}{2}\delta\phi\left(-\bnabla^2+V''(\bphi)-F''(\bphi)\bR\right)\delta\phi
\Biggr]\,.
\eea
As a further simplification we note that defining
\footnote{This change of variables has a trivial Jacobian.}
\be
\sigma''=\sigma'+4\frac{F'(\bar\phi)}{F(\bar\phi)}
\sqrt{\frac{-\bar{\nabla}^2-\frac{\bar R}{3}}{-\bar{\nabla}^2}}\delta\phi\,,
\label{sigmadoubleprime}
\ee 
the gauge fixed Hessian becomes diagonal,
\bea
\label{gfhess}
S^{(2)}_{\mathrm{grav}}&=&\int \mathrm{d}^4x\sqrt{\bg}\,\Biggl[
\frac{1}{4}F(\bphi)\tth_{\mu\nu}\left(-\bnabla^2+\frac{\bR}{6}\right)\tth^{\mu\nu}
-\frac{3}{32}F(\bphi)
\sigma''(-\bnabla^2)\sigma''
\nonumber\\
&+&\frac{1}{2}\delta\phi\,\bigg(-\bnabla^2+V''(\bphi)-F''(\bphi)\bR
+3\frac{F'(\bphi)^2}{F(\bphi)}\left(-\bnabla^2-\frac{\bR}{d-1}\right)\bigg)\,\delta\phi
\Biggr]\,.
\eea
The unimodular physical gauge produces Faddeev-Popov ghost determinants
\begin{equation}
\Delta_{\mathrm{FP}=}\sqrt{{\det}_0(-\bnabla^2)}
\sqrt{{\det}_1\left(-\bnabla^2-\frac{\bR}{4}\right)}\,,
\label{FPdetUnimGauge}
\end{equation}
with the subscripts 0 and 1 denoting the spin of the fields that the corresponding operators act on.
Thus the one-loop partition function reads
\be
Z={\mathrm{e}^{-S_{\mathrm{grav}}[\bar{\phi},\bar{g}]}}\frac{\sqrt{\det\Delta_1}}{\sqrt{\det\Delta_2}
\sqrt{\det\Delta_S}}\,,
\label{pfGR}
\ee
where $\Delta_1=-\bnabla^2-\frac{\bR}{4}$,
$\Delta_0=-\bnabla^2$ and
\be
\Delta_S=-\bnabla^2+E_S\ ,
\qquad
E_S=\frac{F V''-(F^{\prime 2}+F F'')\bR}{F+3F^{\prime 2}}\ .
\label{deltaS}
\ee
This agrees with the standard result for GR with a cosmological constant, 
except for the appearance of the additional scalar determinant.

Consider now the same calculation in UG. The trace fluctuation $h$ is absent from the degrees of freedom from the start and it is therefore not necessary to fix the corresponding gauge. The \textit{SDiff} gauge can be fixed again by setting $\xi'=0$. Altogether this produces the Faddeev-Popov determinant 
\begin{equation}
\Delta^{\mathrm{UG}}_{\mathrm{FP}}=\sqrt{{\det}_1\left(-\bnabla^2-\frac{\bR}{4}\right)}\,.
\end{equation}
On the other hand, as discussed in \cite{Ardon:2017atk,Percacci:2017fsy}, 
the factorization of the volume of \textit{SDiff} produces and additional 
determinant $\sqrt{\det(-\bnabla^2)}$ which cancels the determinant coming from the integration over $\sigma'$, so that the final result is again exactly (\ref{pfGR}). Notably, such an equivalence holds irrespective of the choice of $F(\phi)$. 

In a standard perturbative approach, the beta functions can be read off from the logarithmic divergences.
The one-loop effective action is
\be
\Gamma=S+\frac12\Tr\log\Delta_2-\frac12\Tr\log\Delta_1+
\frac12\Tr\log\Delta_S\,,
\ee
and its divergent parts can be obtained from
\be
\Gamma_{\mathrm{div}}=
-\frac12 \frac{1}{16\pi^2}\log\left(\frac{\Lambda^2}{\mu^2}\right)
\int \mathrm{d}^4x \sqrt{{\bar{g}}}\left[
b_4(\Delta_2)-b_4(\Delta_1)+b_4(\Delta_S)\right]\,,
\label{gammadiv}
\ee
with $\Lambda$ standing for an ultraviolet cutoff and $\mu$ being a reference scale. The first two contributions in \eqref{gammadiv} only give terms of order $R^2$
and are not relevant for the beta functions of interest.
For $\Delta_S$ we have
\bea
-\frac12 \frac{1}{16\pi^2}b_4(\Delta_S)&=&{-\frac12 \frac{1}{16\pi^2}}\left(\frac12 E_S^2-\frac16\bR\, E_S+O(\bR^2)\right)\nonumber\\
&=&
-\frac{1}{64\pi^2}\frac{V''^2}{\left(1+3\frac{F^{\prime2}}{F}\right)}
+
\frac{1}{192\pi^2}\frac{1+6F''+9\frac{F^{\prime2}}{F}}{\left(1+3\frac{F^{\prime2}}{F}\right)}V''\bR+O(\bR^2)\ .
\eea
Inserting (\ref{vpol}) and (\ref{fpol}) and expanding in powers of $\phi$
we obtain the beta functions
\bea
\beta_{\cV}\!\!&=&\!\!
\frac{m^4}{32\pi^2}
\\
\beta_{m^2}\!\!&=&\!\!
\frac{3\lambda}{2\pi^2}m^2
-\frac{6G m^4\xi^2}{\pi}
\\
\beta_\lambda\!\!&=&\!\!
\frac{9\lambda^2}{2\pi^2}
-\frac{72 G m^2 \lambda\xi^2}{\pi}
\\
\beta_{G}\!\!&=&\!\!-\frac{G^2 m^2(1+6\xi)}{6\pi}
\\
\beta_\xi\!\!&=&\!\!
\frac{\lambda(1+6\xi)}{4\pi^2}
+\frac{G m^2\xi^2(1-12\xi)}{\pi}
\eea
We note that the leading terms are the same as for the pure scalar theory, discussed in Appendix \ref{betaFRGap}. Here we only kept correction terms linear in $G$. The explicit one-loop computation reported above leads to the same results in GR and UG, since \eqref{gammadiv} is the same in both cases. Moreover, we have kept only the contributions generated by the 
universal $Q$-functionals. 
The remaining terms are associated to power divergences and are not universal.
The conclusion of this explicit calculation agrees with our statement in Sec.~\ref{Sect:EPI}. In particular, the non-minimal scalar-gravity coupling does not change this conclusion. 

As is well-known, quantum-gravity contributions to matter beta functions can be gauge dependent. 
The ``unimodular physical gauge'' 
can be obtained from the standard two-parameter linear covariant gauge condition for Diff-invariant theories, namely
\begin{equation}
\bar{\nabla}^{\nu}{h_{\nu\mu}} - \frac{1+\beta}{4}\bar{\nabla}_{\mu}h = 
\alpha b_\mu\,,
\label{gaugecondition1}
\end{equation}
with $b_\mu$ being a fixed function, by taking the limits $\alpha\to 0$ and $\beta\to -\infty$.  
For generic $\alpha$, $\beta$, the beta functions $\beta_\cV$ and $\beta_G$ are left unchanged, while the others become,
\begin{eqnarray}
\beta_{m^2} &=& \frac{3m^2 \lambda}{2\pi^2}+\frac{2Gm^4(4\alpha-3(2+(3-\beta)\xi)^2)}{(3-\beta)^2\pi}\,,\nonumber\\
\beta_\lambda &=& \frac{9\lambda^2}{2\pi^2}-\frac{8Gm^2 \lambda (12-4\alpha+24(3-\beta)\xi+9(3-\beta)^2\xi^2)}{(3-\beta)^2\pi}\,,\nonumber\\
\beta_\xi &=& \frac{\lambda (1+6\xi)}{4\pi^2}-\frac{Gm^2}{12\pi}\EuScript{F}(\alpha,\beta,\xi)\,.
\label{betasgeneric}
\end{eqnarray}
The contribution $\EuScript{F}(\alpha,\beta,\xi)$ is lengthy and collected in the Appendix~\ref{appB}. In the limit $\beta\to -\infty$, the beta functions turn out to be $\alpha$-independent.
It is also worth mentioning that the first two of these beta functions
are also independent of another parameter that can be introduced
in the definition of the measure, namely the use of a densitized
metric as a quantum field, see, e.g., \cite{Ohta:2016npm}.

As a side comment, at one-loop order, the universal gravitational correction to the quartic coupling $\lambda$ at vanishing non-minimal coupling is negative, irrespective of the values of the gauge parameter $\beta$, provided that\footnote{In an Euclidean setting, $\alpha$ has to be non-negative.} $\alpha < 3$,
\begin{equation}
\beta_\lambda\Big|_{\mathrm{grav}} = -\lambda\frac{32Gm^2}{\pi}\frac{3-\alpha}{(3-\beta)^2}\,.
\label{betalambdaxizero}
\end{equation}
At $\alpha = 3$ or $\beta\to\pm\infty$, the contribution vanishes at one loop. In particular, in the unimodular physical gauge, the gravitational contribution vanishes at vanishing $\xi$. However, in such a gauge, if the non-minimal coupling is included, the contribution is always negative 
at leading order in $G$. Hence, such a contribution can balance the non-gravitational contribution to the one-loop running of $\lambda$ - which is 
positive and leads to the well-known triviality problem. 
In order to circumvent the issues due to the gauge dependence, 
and of the non-universal power-law terms,
one will have to to compute a gauge invariant physical observable
possibly along the lines of \cite{Frob:2017lnt}.

\subsection{A universal beta function?}

In \cite{Herrero-Valea:2020xaq}, it was argued that the dimensionless combination of couplings
\begin{equation}
\Delta = \frac{(Gm^2)^2}{\lambda}\,,
\label{deltadef}
\end{equation}
has a universal beta function and carries a physical meaning. By quantizing UG in the presence of non-minimally coupled scalar fields, the authors claim that the results differ in GR and UG. 
More precisely, taking into account the differences in notation,
their result for UG is
\be
\beta_\Delta^\textmd{UG}=\Delta\frac{-9\lambda+2\pi G m^2
(-4-24\xi+180\xi^2)}{6\pi^2}
\ee
while their result for GR is
\be
\beta_\Delta^\textmd{GR}=\Delta\frac{-9\lambda+2\pi G m^2
(-4+156\xi+180\xi^2)}{6\pi^2}\ .
\ee
Hence, $\Delta$ would be a physical quantity able to distinguish GR and UG if the scalar field is non-minimally coupled to the gravitational field. 

Using our previous calculations, we cannot distinguish UG and GR non-minimally coupled to scalars at one-loop simply because the path integrals are the same. In particular, in the unimodular physical gauge, we obtain
\begin{equation}
\beta_\Delta = \Delta\frac{-9\lambda + 2Gm^2\pi (-1-6\xi+180\xi^2)}{6\pi^2}\ ,
\label{betadeltaexpparamunigauge}
\end{equation}
which differs from either of the results above.
These discrepancies may be ascribed to the fact that we are using a different parameterizarion of the metric and a different implementation of the 
unimodularity condition. 

What is perhaps more important is that, even if we stick to our computation scheme, the quantity $\Delta$ is gauge dependent.
In fact, in the linear covariant gauge \eqref{gaugecondition1},
the result is
\begin{equation}
\beta_{\Delta} = \frac{\Delta (-9(3-\beta)^2\lambda-2Gm^2\pi (48\alpha + \beta^2 \EuScript{A}_1 (\xi)+6\beta \EuScript{A}_2 (\xi)-27\EuScript{A}_3 (\xi)))}{6\pi^2(3-\beta)^2}\,,
\label{betadeltaexpgeneralgauge}
\end{equation}
with
\be
\EuScript{A}_1 (\xi)=1+6\xi-180\xi^2\ ,\quad
\EuScript{A}_2 (\xi)=-1+66\xi+180\xi^2\ ,\quad
\EuScript{A}_3 (\xi)=5+46\xi+60\xi^2\,.
\label{exprbetadeltageneralgauge}
\ee
Thus, even in the absence of $\xi$, the beta function of $\Delta$ is gauge dependent and comparing results for GR and UG would be problematic. We also remark that, in the limit $\beta\to\pm\infty$, eq.\eqref{betadeltaexpgeneralgauge} reduces to \eqref{betadeltaexpparamunigauge} irrespective of $\alpha$.

\subsection{Dynamical gravitons: GR in linear parametrization}

So far, the explicit one-loop computations were performed using the exponential parameterization of the metric. 
In this parameterization, 
the unimodularity condition simply amounts to removing 
the trace mode of the gravitational field fluctuation $h_{\mu\nu}$. While field redefinitons, properly done, 
should not affect the result of physical quantities, 
there are several subtleties when changing from one parameterization to another in quantum gravity. In this section, we present the one-loop results for the scalar-gravity system with a non-minimal coupling in the so-called linear parameterization, i.e., 
\begin{equation}
g_{\mu\nu} = \bar{g}_{\mu\nu} + h_{\mu\nu}\,,
\label{linearparam1gr}
\end{equation} 
in the linear covariant gauges \eqref{gaugecondition1}. This system was studied, e.g., in \cite{Narain:2009fy}, but the beta functions were computed with the functional renormalization group and contained also non-universal terms. Here, we select just the universal contributions, that are related to logarithmic divergences. 
In a general gauge $(\alpha,\beta)$, the result is completely equivalent to \eqref{betasgeneric} apart from the beta function of the non-minimal coupling $\beta_\xi$ which reads
\begin{equation}
\beta_\xi = \frac{\lambda(1+6\xi)}{4\pi^2}-\frac{Gm^2}{12\pi}\EuScript{G}(\alpha,\beta,\xi)\,,
\label{betaxilinparam}
\end{equation}
where the explicit expression for $\EuScript{G}(\alpha,\beta,\xi)$ is reported in Appendix~\ref{appB}. In particular, if we take $\alpha \to 0$ and $\beta\to\pm\infty$, we obtain
\begin{equation}
\EuScript{G}(0,\pm\infty,\xi)= 6(-13+10\xi^2+24\xi^3)\,,
\label{galpha0betainftylinearparam}
\end{equation}
which differs from $\EuScript{F}(0,\pm\infty,\xi)$. The beta function of $\Delta$ in a general linear covariant gauge and in linear parameterization is the same as \eqref{betadeltaexpgeneralgauge}. Hence, although gauge-dependent, $\Delta$ seems to display some kind of universality as far as 
different choices of parameterization is concerned. This fact is not very 
surprising given that, the only beta function in the linear parameterizarion that differs from the exponential parameterization at one-loop is $\beta_\xi$ and it does not enter the definition of $\beta_\Delta$.

In \cite{Steinwachs:2011zs}, Kamenshchick and Steinwachs 
(see, also, \cite{Shapiro:1995yc})
investigated the one-loop divergences of a more general theory 
than the one considered in this work. In particular, they have considered 
a scalar-gravity action $S_{\mathrm{KS}}[g,\Phi]$ expressed as\footnote{The functions $U$, $G$, $V$ of
\cite{Steinwachs:2011zs} correspond to $-F$, $-1$, $-V$ in our notation.}
\begin{equation}
S_{\mathrm{KS}}[g,\Phi] = \int\mathrm{d}^4x\sqrt{g}\,\bigg(V(\tilde{\Phi})-F(\tilde{\Phi}) R + \frac{1}{2}g^{\mu\nu}G(\tilde{\Phi})\nabla_\mu \Phi^a \nabla_\nu \Phi_a\bigg)\,,
\label{steinwachsk1}
\end{equation}
where $a=1,\ldots ,N$ and $N$ is a positive integer. The functions $V$ and $F$ depend on
$\tilde{\Phi} = \sqrt{\delta_{ab}\Phi^a \Phi^b}$.
The gauge condition used in \cite{Steinwachs:2011zs} is
\begin{equation}
F_{\mu} = \sqrt{F(\tilde{\Phi})}\bigg(\bar{\nabla}^{\alpha}h_{\alpha\mu}-\frac{1}{2}\bar{\nabla}_\mu h - \frac{F^\prime (\tilde{\Phi})}{F(\tilde{\Phi})}n^a\bar{\nabla}_{\mu}\varphi_a\bigg)\,,
\label{steinwachsk3}
\end{equation}
with $\varphi_a$ the scalar field fluctuations and $n^a = \Phi^a/\tilde{\Phi}$. 
Unfortunately, our gauge condition \eqref{gaugecondition1} is not deformable to this, and therefore
we cannot directly compare our results with theirs.
However, we can extract from their work
the beta function of $\Delta$ in the gauge \eqref{steinwachsk3}.
The authors employed the linear parameterization of the metric and the reduction to the a single-scalar non-minimally coupled to gravity is achieved by taking $N\to1$, $\Phi^a = n^a =1$ and $\tilde{\Phi}\to\phi$. Moreover, in order to have the same scalar-tensor action we discussed in this work, one has to take
$G(\tilde{\Phi}) \to 1$
and their quantity $s$ has to be identified as
\begin{equation}
s=-\frac{F}{F+3F^{\prime 2}}
=\frac{-1-8\pi \xi G\phi^2}{1+8\pi \xi G\phi^2(1+6\xi)}\,.
\end{equation}
The beta functionals of $V$ and $F$ are
$\beta_V=2\alpha_1$, $\beta_F=2\alpha_2$,
where $\alpha_1$ and $\alpha_2$ are given in 
their equations (48) and (49).
From there we read off
\bea
\beta_{m^2}\!\!&=&\!\!
\frac{3\lambda}{2\pi^2}m^2
-\frac{2G m^4(2+4\xi+3\xi^2)}{\pi}\,,
\\
\beta_\lambda\!\!&=&\!\!
\frac{9\lambda^2}{2\pi^2}
-\frac{8G m^2 \lambda(2+8\xi+9\xi^2)}{\pi}\,,
\\
\beta_{G}\!\!&=&\!\!-\frac{G^2 m^2(1+6\xi)}{6\pi}\,,
\\
\beta_\xi\!\!&=&\!\!
\frac{\lambda(1+6\xi)}{4\pi^2}-\frac{G m^2(13-16\xi-39\xi^2-36\xi^3)}{3\pi}\ ,
\eea
which gives
\be
\beta_\Delta=\Delta\frac{-9\lambda
+10\pi G m^2
(5+30\xi+36\xi^2)}{6\pi^2}\,.
\ee
This confirms once more that the first and last term in the
fraction are universal, but not the other ones.

\section{Conclusions \label{Concl}}

Disregarding the single global spacetime volume degree of freedom,
we have shown at a formal path integral level that the classical
equivalence between general (\textit{Diff}--invariant)
and unimodular (\textit{SDiff}--invariant)
versions of gravity theories,
can be maintained at the quantum level\footnote{The equivalence of GR and UG
in the presence of an independent connection,
deserves a separate investigation, which is ongoing.}.
This is true independently of the choice of the action
and also in the presence of matter.

We have then given an explicit one-loop example of this,
by computing the universal parts of the beta functions of 
scalars coupled to gravity.
In spite of significant differences in the two cases,
the beta functions turn out to be the same.
We have then compared these results to those of 
\cite{Herrero-Valea:2020xaq},
who also made the same comparison.
Our beta function for the dimensionless combination $\Delta$ 
differs from theirs both for GR and UG.
The differences can probably to be ascribed
at least in part to the different way they implement unimodularity.
A more detailed analysis has shown that the beta function
of $\Delta$ is actually gauge-dependent, 
so that it is not a sufficiently good test.
There are two terms in the beta function of $\Delta$
that are the same in all gauges and are the same across
all calculations we could find in the literature, 
whereas other terms have strong gauge dependence.
For the future, it will be important to identify a genuinely
universal combination of couplings, or another observable that
can act as a benchmark.

We conclude with some comments on the cosmological constant. 
In UG, a ``cosmological term'' 
$\frac{\Lambda}{8\pi G}\int \mathrm{d}^4x\,\omega$ 
in the Lagrangian is just an additive, field-independent term
that does not affect the equations of motion
and can be absorbed in the overall normalization of
the functional integral. Thus, it has no physical effect.
GR is only (classically) equivalent to UG if we
impose that the total volume of spacetime is fixed.
In this restricted theory the cosmological term 
in (\ref{volconstr}) is a Lagrange multiplier,
whose value is ultimately related to the volume
through the equations of motion.

Computations of the beta functions performed in the so-called unimodular gauge \cite{Percacci:2015wwa} show that the 
cosmological constant decouples from the system of beta functions.
This resembles simpler calculations involving the functional renormalization group, 
where a field-independent contribution is generated by the flow and can be cancelled by a proper normalization of the vacuum energy. 
This suggests that its quartic running is unphysical.
This is in line with other hints coming from
different directions \cite{Pagani:2019vfm,Pagani:2020say,Becker:2020mjl,Donoghue:2020hoh}. 
This and related issues deserve to be investigated further.

\section*{Acknowledgments}

We thank Mario Herrero-Valea, Christian Steinwachs, Reinhard Alkofer
and Kevin Falls for valuable discussions and comments. 
ADP acknowledges CNPq under the grant PQ-2 (309781/2019-1), FAPERJ under the “Jovem Cientista do Nosso Estado” program (E26/202.800/2019), and NWO under the VENI Grant (VI.Veni.192.109) for financial support.

\appendix

\section{Derivation of beta functions from the Functional Renormalization 
Group} \label{betaFRGap}

\subsection{Extracting the universal terms from the FRG}
\label{uniFRG}

In the literature, the beta functions of gravity, with or without matter, 
have been often calculated in the Functional Renormalization Group (FRG) framework, see, e.g., \cite{Percacci:2017fkn,Dupuis:2020fhh,Pawlowski:2005xe,Gies:2006wv} for reviews on the subject).
Since the FRG is based on a momentum cutoff,
the beta functions contain terms proportional to powers of the
cutoff, that are not seen with other techniques.
In this appendix we discuss the way in which one can recover
from the FRG the standard one loop beta functions that one would
see, e.g. in dimensional regularization.
For a more detailed discussion of the relation between the FRG
and dimensional regularization we refer to \cite{Baldazzi:2020vxk}.

In the FRG, a cutoff function $R_{k}$ is introduced by hand in the
quadratic part of the action,
in order to suppress the contribution to the functional integral 
of modes with (Euclidean) momenta smaller than a cutoff scale $k$. 
This leads to a coarse-grained effective action $\Gamma_k$ which coincides with the full effective action at $k=0$.
The flowing action $\Gamma_k$ obeys the flow equation
\begin{equation}
k\frac{\mathrm{d}}{\mathrm{d}k}\Gamma_k \equiv \partial_t \Gamma_k = \frac{1}{2}\mathrm{Tr}\left[(\Gamma^{(2)}_k+R_k)^{-1}\partial_t R_k\right]\ 
,
\label{apa1}
\end{equation}
where $\Gamma^{(2)}_k$ is the Hessian constructed from $\Gamma_k$.
For our present purposes it will be enough to consider the 
simple case of scalar fields in a background metric, 
with a Hessian of the form $\Gamma^{(2)}_k=\Delta$
where $\Delta$ is a Laplace-type operator:
\be
\Delta=-\nabla^2+E\ ;
\quad E=m^2+12\lambda\phi^2-\xi R\ .
\label{comparison2}
\ee
This derives from a scalar action
containing a potential and a non-minimal coupling to gravity.
Then the r.h.s. of the flow equation is a function $W(\Delta)$
that, for constant $\phi$, can be evaluated as
\begin{eqnarray}
\mathrm{Tr}W(\Delta) &=& \frac{1}{(4\pi)^{d/2}}\Big[Q_{d/2}(W)B_0 (\Delta)+Q_{d/2 - 1}(W)B_2 (\Delta)+\ldots+Q_0 (W)B_d (\Delta)
+\ldots\Big]\,,
\label{apa3}
\end{eqnarray}
with the $Q$-functionals defined as
\begin{equation}
Q_n (W) = \frac{(-1)^{k}}{\Gamma (n+k)}\int^{\infty}_0 \mathrm{d}z\, z^{n+k-1}W^{(k)}(z)\,.
\label{apa4}
\end{equation}
In eq.\eqref{apa4}, $n\in \mathbf{R}$, $W^{(k)}(z)$ stands for the $k$-th 
derivative of $W$ with respect to $z$. If $n>0$, then $k=0$.
Otherwise, $k$ is a positive integer such that $n+k>0$. 
The heat kernel coefficients are
$B_n (\Delta) = \int\mathrm{d}^dx\sqrt{g}~\mathrm{Tr}~b_n (\Delta)$,
where 
\begin{eqnarray}
b_0 &=& 1\,,\qquad\quad
b_2=\frac{R}{6}-E\,,
\nonumber\\
b_4 &=& \frac{1}{180}\left(R^{\mu\nu\alpha\beta}R_{\mu\nu\alpha\beta}-R^{\mu\nu}R_{\mu\nu}+\frac{5}{2}R^2
\right)
-\frac{1}{6}R E+\frac{1}{2}E^2 \,.
\label{apa10}
\end{eqnarray}
In the flow equation, we are interested in computing $Q$-functionals 
of the form
\begin{equation}
W(z) = \frac{\partial_t R_k (z)}{(P_k (z) )^{m}}\,,
\label{apa6}
\end{equation}
where $P_k (z) = z+R_k (z)$. If, $m = n+1$, then one can show that
\begin{equation}
Q_{n}\left(\frac{\partial_t R_k}{P^{n+1}_k}\right) = \frac{2}{\Gamma (n+1)}\,,
\label{apa7}
\end{equation}
is ``universal'', i.e. independent of the shape of $R_k$. 
For certain cutoff schemes the denominator in the function $W$
is $P_k+E$, and
\begin{equation}
Q_{n}\left(W\right)=Q_{n}\left(\frac{\partial_t R_k}{(P_k+E)^m}\right)\,,
\label{apa8}
\end{equation}
are, in general, non-universal quantities. Nevertheless, one can extract universal parts of each $Q$-functional defined in eq.\eqref{apa8} by expanding in $E$:
\begin{equation}
Q_n\left(\frac{\partial_t R_k}{(P_k + E)^m}\right) = Q_n\left(\frac{\partial_t R_k}{P^m_k}\left(1-m\frac{E}{P_k}+\frac{m(m+1)}{2}\frac{E^2}{P^{2}_k}-\frac{m(m+1)(m+2)}{3!}\frac{E^3}{P^3_k}+\ldots\right)\right)\,,
\label{apa9}
\end{equation}
and exploiting the linearity of the $Q$-functionals
to  pick up the contribution which satisfies $n=m+1$.

Consider first a ``type III'' cutoff, see, e.g., \cite{Codello:2008vh,Percacci:2017fkn}. The beta functional is 
\be
\dot\Gamma_k=\frac{1}{32\pi^2}\Bigg[Q_2\left(\frac{\dot R_k}{P_k}\right)B_0(\Delta)+Q_1\left(\frac{\dot R_k}{P_k}\right)B_2(\Delta)+Q_0\left(\frac{\dot R_k}{P_k}\right)B_4(\Delta)+\ldots\Bigg]\,.
\label{comparison3}
\ee
Only the last term is universal.
Thus
\be
\dot\Gamma_k\Big|_{\mathrm{univ}}=\frac{2}{32\pi^2}\int \mathrm{d}^4x\sqrt{g}\,b_4(\Delta)\,.
\label{comparison4}
\ee
The relevant terms (up to linear order in $R$ which are not total derivatives) are
\bea
b_4&\sim& \frac12 E^2-\frac16 R \, E
\nonumber\\
&\sim&\frac12 m^4+72\lambda^2\phi^4
+12\lambda m^2\phi^2
+\left(\xi+\frac16\right) m^2 R
+2\lambda(6\xi+1)\phi^2 R\,.
\label{comparison5}
\eea
From here one reads off the beta functions
\bea
\beta_{\cV}\!\!&=&\!\!
\frac{m^4}{32\pi^2}\,,
\\
\beta_{m^2}\!\!&=&\!\!
\frac{3\lambda m^2}{2\pi^2}\,,
\\
\beta_\lambda\!\!&=&\!\!
\frac{9\lambda^2}{2\pi^2}\,,
\\
\beta_{Z_N}\!\!&=&\!\!\frac{1+6\xi}{96\pi^2}m^2\,,
\\
\beta_\xi\!\!&=&\!\!
\frac{\lambda(1+6\xi)}{4\pi^2}\,.
\label{comparison6}
\eea

The same result can be obtained in a more laborious way using a 
``type I'' cutoff. In this case
\be
\dot\Gamma_k\!\!=\!\!\frac{1}{32\pi^2}
\Bigg[Q_2\left(\frac{\dot R_k}{P_k+E}\right)B_0(-\nabla^2)
+Q_1\left(\frac{\dot R_k}{P_k+E}\right)B_2(-\nabla^2)
+Q_0\left(\frac{\dot R_k}{P_k+E}\right)
B_4(-\nabla^2)
+\ldots\Bigg]\,.
\label{comparison7}
\ee
The universal terms come from all three pieces in this expression, when one expands in $E$: the third term in the expansion for $Q_2$, the second for $Q_1$ and the leading term for $Q_0$. In the latter term, $B_4(-\nabla^2)$ is of order $R^2$ and does not concern us. The rest is
\bea
\dot\Gamma_k\!\!&\sim&\!\!\frac{1}{32\pi^2}
\Bigg[Q_2\left(\frac{\dot R_k}{P_k^3}\right) E^2B_0(-\nabla^2)
+Q_1\left(\frac{\dot R_k}{P_k^2}\right)(-E)B_2(-\nabla^2)
+\ldots\Bigg]\,,
\nonumber\\
&=&\frac{1}{32\pi^2}\int \mathrm{d}^4x\sqrt{g}
\Bigg[E^2-2E \frac16 R
+\ldots\Bigg]\,,
\label{comparison8}
\eea
which is clearly the same as before, and therefore leads to the same beta 
functions. Note that the universal parts of the beta functions are those that come from the dimensionless $Q$-functionals and therefore are independent of $k$, see eq.\ref{apa7}.

\section{Some General Expressions}\label{appB}

In this appendix, we collect some long expressions that were omitted in the main text. In particular, the beta function of the non-minimal coupling $\xi$ depends on the choice of the metric parameterization. Hence, in the exponential parameterization in a general linear covariant gauge \eqref{gaugecondition1}, the factor $\EuScript{F}$ in eq.\eqref{betasgeneric} is
\begin{equation}
\EuScript{F}(\alpha,\beta,\xi)=-4\frac{\EuScript{F}_1 (\alpha,\beta,\xi)-3 \left(\EuScript{F}_2 (\beta,\xi)+\EuScript{F}_3 (\beta,\xi)+\EuScript{F}_4 (\beta,\xi)+\EuScript{F}_5 (\beta,\xi)+\EuScript{F}_6 (\beta,\xi)\right)}{(3-\beta)^4}\,,
\label{f}
\end{equation}
with
\begin{eqnarray}
\EuScript{F}_1 (\alpha,\beta,\xi) &=& 24 \alpha ^2+2 \alpha  (\beta ^2 (24 \xi +1)-18 \beta  (4 \xi +1)-27)\,,\nonumber\\
\EuScript{F}_2 (\beta,\xi) &=& \beta ^4 \xi ^2 (12 \xi -1)\,,\nonumber\\
\EuScript{F}_3 (\beta,\xi) &=& -4 \beta ^3 \xi  \left(36 \xi ^2+9 \xi -1\right)\,,\nonumber\\
\EuScript{F}_4 (\beta,\xi) &=& \beta ^2 \left(648 \xi ^3+342 \xi ^2+36 \xi -2\right)\,,\nonumber\\
\EuScript{F}_5 (\beta,\xi) &=& -12 \beta  \left(108 \xi ^3+81 \xi ^2+15 
\xi +1\right)\,,\nonumber\\
\EuScript{F}_6 (\beta,\xi) &=& 9 \left(108 \xi ^3+99 \xi ^2+12 \xi -2\right)\,.
\label{fs}
\end{eqnarray}
As for the linear parameterization, the expression for $\EuScript{G}(\alpha,\beta,\xi)$ in \eqref{betaxilinparam} is
\begin{equation}
\EuScript{G}(\alpha,\beta,\xi)=2\frac{\EuScript{G}_{1}(\alpha,\beta)+\EuScript{G}_{2}(\alpha,\beta,\xi)+3\EuScript{G}_4 (\beta,\xi)}{(3-\beta)^4}
\label{g}
\end{equation}
with
\begin{eqnarray}
\EuScript{G}_{1}(\alpha,\beta) &=& -3 \alpha ^2 (3 (\beta -6) \beta  ((\beta -6) \beta +18)+259)\,,\nonumber\\
\EuScript{G}_{2}(\alpha,\beta,\xi) &=& \alpha  (2 \beta  (\beta  (3 (\beta -12) \beta +24 \xi +194)-396)-432 \xi +630)\,,\nonumber\\
\EuScript{G}_3(\beta,\xi)&=& -13 \beta ^4+112 \beta ^3-458 \beta ^2+24 (\beta -3)^4 \xi ^3\,,\nonumber\\
\EuScript{G}_4(\beta,\xi)&=& \left(2 (\beta  (5 \beta -42)+117) (\beta -3)^2 \xi ^2-8 (\beta  (5 \beta -12)+27) (\beta -3) \xi +888 \beta -657\right)\,.\nonumber\\
\label{gs}
\end{eqnarray}

\bibliography{refs}

\begin{thebibliography}{65}%
\makeatletter
\providecommand \@ifxundefined [1]{%
 \@ifx{#1\undefined}
}%
\providecommand \@ifnum [1]{%
 \ifnum #1\expandafter \@firstoftwo
 \else \expandafter \@secondoftwo
 \fi
}%
\providecommand \@ifx [1]{%
 \ifx #1\expandafter \@firstoftwo
 \else \expandafter \@secondoftwo
 \fi
}%
\providecommand \natexlab [1]{#1}%
\providecommand \enquote  [1]{``#1''}%
\providecommand \bibnamefont  [1]{#1}%
\providecommand \bibfnamefont [1]{#1}%
\providecommand \citenamefont [1]{#1}%
\providecommand \href@noop [0]{\@secondoftwo}%
\providecommand \href [0]{\begingroup \@sanitize@url \@href}%
\providecommand \@href[1]{\@@startlink{#1}\@@href}%
\providecommand \@@href[1]{\endgroup#1\@@endlink}%
\providecommand \@sanitize@url [0]{\catcode `\\12\catcode `\$12\catcode
  `\&12\catcode `\#12\catcode `\^12\catcode `\_12\catcode `\%12\relax}%
\providecommand \@@startlink[1]{}%
\providecommand \@@endlink[0]{}%
\providecommand \url  [0]{\begingroup\@sanitize@url \@url }%
\providecommand \@url [1]{\endgroup\@href {#1}{\urlprefix }}%
\providecommand \urlprefix  [0]{URL }%
\providecommand \Eprint [0]{\href }%
\providecommand \doibase [0]{https://doi.org/}%
\providecommand \selectlanguage [0]{\@gobble}%
\providecommand \bibinfo  [0]{\@secondoftwo}%
\providecommand \bibfield  [0]{\@secondoftwo}%
\providecommand \translation [1]{[#1]}%
\providecommand \BibitemOpen [0]{}%
\providecommand \bibitemStop [0]{}%
\providecommand \bibitemNoStop [0]{.\EOS\space}%
\providecommand \EOS [0]{\spacefactor3000\relax}%
\providecommand \BibitemShut  [1]{\csname bibitem#1\endcsname}%
\let\auto@bib@innerbib\@empty
\bibitem [{\citenamefont {Anderson}\ and\ \citenamefont
  {Finkelstein}(1971)}]{Anderson:1971pn}%
  \BibitemOpen
  \bibfield  {author} {\bibinfo {author} {\bibfnamefont {J.~L.}\ \bibnamefont
  {Anderson}}\ and\ \bibinfo {author} {\bibfnamefont {D.}~\bibnamefont
  {Finkelstein}},\ }\bibfield  {title} {\bibinfo {title} {{Cosmological
  constant and fundamental length}},\ }\href
  {https://doi.org/10.1119/1.1986321} {\bibfield  {journal} {\bibinfo
  {journal} {Am. J. Phys.}\ }\textbf {\bibinfo {volume} {39}},\ \bibinfo
  {pages} {901} (\bibinfo {year} {1971})}\BibitemShut {NoStop}%
\bibitem [{\citenamefont {van~der Bij}\ \emph {et~al.}(1982)\citenamefont
  {van~der Bij}, \citenamefont {van Dam},\ and\ \citenamefont
  {Ng}}]{vanderBij:1981ym}%
  \BibitemOpen
  \bibfield  {author} {\bibinfo {author} {\bibfnamefont {J.~J.}\ \bibnamefont
  {van~der Bij}}, \bibinfo {author} {\bibfnamefont {H.}~\bibnamefont {van
  Dam}},\ and\ \bibinfo {author} {\bibfnamefont {Y.~J.}\ \bibnamefont {Ng}},\
  }\bibfield  {title} {\bibinfo {title} {{The Exchange of Massless Spin Two
  Particles}},\ }\href {https://doi.org/10.1016/0378-4371(82)90247-3}
  {\bibfield  {journal} {\bibinfo  {journal} {Physica A}\ }\textbf {\bibinfo
  {volume} {116}},\ \bibinfo {pages} {307} (\bibinfo {year}
  {1982})}\BibitemShut {NoStop}%
\bibitem [{\citenamefont {Buchmuller}\ and\ \citenamefont
  {Dragon}(1988)}]{Buchmuller:1988wx}%
  \BibitemOpen
  \bibfield  {author} {\bibinfo {author} {\bibfnamefont {W.}~\bibnamefont
  {Buchmuller}}\ and\ \bibinfo {author} {\bibfnamefont {N.}~\bibnamefont
  {Dragon}},\ }\bibfield  {title} {\bibinfo {title} {{Einstein Gravity From
  Restricted Coordinate Invariance}},\ }\href
  {https://doi.org/10.1016/0370-2693(88)90577-1} {\bibfield  {journal}
  {\bibinfo  {journal} {Phys. Lett. B}\ }\textbf {\bibinfo {volume} {207}},\
  \bibinfo {pages} {292} (\bibinfo {year} {1988})}\BibitemShut {NoStop}%
\bibitem [{\citenamefont {Buchmuller}\ and\ \citenamefont
  {Dragon}(1989)}]{Buchmuller:1988yn}%
  \BibitemOpen
  \bibfield  {author} {\bibinfo {author} {\bibfnamefont {W.}~\bibnamefont
  {Buchmuller}}\ and\ \bibinfo {author} {\bibfnamefont {N.}~\bibnamefont
  {Dragon}},\ }\bibfield  {title} {\bibinfo {title} {{Gauge Fixing and the
  Cosmological Constant}},\ }\href
  {https://doi.org/10.1016/0370-2693(89)91608-0} {\bibfield  {journal}
  {\bibinfo  {journal} {Phys. Lett. B}\ }\textbf {\bibinfo {volume} {223}},\
  \bibinfo {pages} {313} (\bibinfo {year} {1989})}\BibitemShut {NoStop}%
\bibitem [{\citenamefont {Weinberg}(1989)}]{Weinberg:1988cp}%
  \BibitemOpen
  \bibfield  {author} {\bibinfo {author} {\bibfnamefont {S.}~\bibnamefont
  {Weinberg}},\ }\bibfield  {title} {\bibinfo {title} {{The Cosmological
  Constant Problem}},\ }\href {https://doi.org/10.1103/RevModPhys.61.1}
  {\bibfield  {journal} {\bibinfo  {journal} {Rev. Mod. Phys.}\ }\textbf
  {\bibinfo {volume} {61}},\ \bibinfo {pages} {1} (\bibinfo {year}
  {1989})}\BibitemShut {NoStop}%
\bibitem [{\citenamefont {Unruh}(1989)}]{Unruh:1988in}%
  \BibitemOpen
  \bibfield  {author} {\bibinfo {author} {\bibfnamefont {W.~G.}\ \bibnamefont
  {Unruh}},\ }\bibfield  {title} {\bibinfo {title} {{A Unimodular Theory of
  Canonical Quantum Gravity}},\ }\href
  {https://doi.org/10.1103/PhysRevD.40.1048} {\bibfield  {journal} {\bibinfo
  {journal} {Phys. Rev. D}\ }\textbf {\bibinfo {volume} {40}},\ \bibinfo
  {pages} {1048} (\bibinfo {year} {1989})}\BibitemShut {NoStop}%
\bibitem [{\citenamefont {Unruh}\ and\ \citenamefont
  {Wald}(1989)}]{Unruh:1989db}%
  \BibitemOpen
  \bibfield  {author} {\bibinfo {author} {\bibfnamefont {W.~G.}\ \bibnamefont
  {Unruh}}\ and\ \bibinfo {author} {\bibfnamefont {R.~M.}\ \bibnamefont
  {Wald}},\ }\bibfield  {title} {\bibinfo {title} {{Time and the Interpretation
  of Canonical Quantum Gravity}},\ }\href
  {https://doi.org/10.1103/PhysRevD.40.2598} {\bibfield  {journal} {\bibinfo
  {journal} {Phys. Rev. D}\ }\textbf {\bibinfo {volume} {40}},\ \bibinfo
  {pages} {2598} (\bibinfo {year} {1989})}\BibitemShut {NoStop}%
\bibitem [{\citenamefont {Henneaux}\ and\ \citenamefont
  {Teitelboim}(1989)}]{Henneaux:1989zc}%
  \BibitemOpen
  \bibfield  {author} {\bibinfo {author} {\bibfnamefont {M.}~\bibnamefont
  {Henneaux}}\ and\ \bibinfo {author} {\bibfnamefont {C.}~\bibnamefont
  {Teitelboim}},\ }\bibfield  {title} {\bibinfo {title} {{The Cosmological
  Constant and General Covariance}},\ }\href
  {https://doi.org/10.1016/0370-2693(89)91251-3} {\bibfield  {journal}
  {\bibinfo  {journal} {Phys. Lett. B}\ }\textbf {\bibinfo {volume} {222}},\
  \bibinfo {pages} {195} (\bibinfo {year} {1989})}\BibitemShut {NoStop}%
\bibitem [{\citenamefont {Ellis}\ \emph {et~al.}(2011)\citenamefont {Ellis},
  \citenamefont {van Elst}, \citenamefont {Murugan},\ and\ \citenamefont
  {Uzan}}]{Ellis:2010uc}%
  \BibitemOpen
  \bibfield  {author} {\bibinfo {author} {\bibfnamefont {G.~F.~R.}\
  \bibnamefont {Ellis}}, \bibinfo {author} {\bibfnamefont {H.}~\bibnamefont
  {van Elst}}, \bibinfo {author} {\bibfnamefont {J.}~\bibnamefont {Murugan}},\
  and\ \bibinfo {author} {\bibfnamefont {J.-P.}\ \bibnamefont {Uzan}},\
  }\bibfield  {title} {\bibinfo {title} {{On the Trace-Free Einstein Equations
  as a Viable Alternative to General Relativity}},\ }\href
  {https://doi.org/10.1088/0264-9381/28/22/225007} {\bibfield  {journal}
  {\bibinfo  {journal} {Class. Quant. Grav.}\ }\textbf {\bibinfo {volume}
  {28}},\ \bibinfo {pages} {225007} (\bibinfo {year} {2011})},\ \Eprint
  {https://arxiv.org/abs/1008.1196} {arXiv:1008.1196 [gr-qc]} \BibitemShut
  {NoStop}%
\bibitem [{\citenamefont {Ellis}(2014)}]{Ellis:2013uxa}%
  \BibitemOpen
  \bibfield  {author} {\bibinfo {author} {\bibfnamefont {G.~F.~R.}\
  \bibnamefont {Ellis}},\ }\bibfield  {title} {\bibinfo {title} {{The
  Trace-Free Einstein Equations and inflation}},\ }\href
  {https://doi.org/10.1007/s10714-013-1619-5} {\bibfield  {journal} {\bibinfo
  {journal} {Gen. Rel. Grav.}\ }\textbf {\bibinfo {volume} {46}},\ \bibinfo
  {pages} {1619} (\bibinfo {year} {2014})},\ \Eprint
  {https://arxiv.org/abs/1306.3021} {arXiv:1306.3021 [gr-qc]} \BibitemShut
  {NoStop}%
\bibitem [{\citenamefont {Wilczek}(1984)}]{Wilczek:1983as}%
  \BibitemOpen
  \bibfield  {author} {\bibinfo {author} {\bibfnamefont {F.}~\bibnamefont
  {Wilczek}},\ }\bibfield  {title} {\bibinfo {title} {{Foundations and Working
  Pictures in Microphysical Cosmology}},\ }\href
  {https://doi.org/10.1016/0370-1573(84)90206-0} {\bibfield  {journal}
  {\bibinfo  {journal} {Phys. Rept.}\ }\textbf {\bibinfo {volume} {104}},\
  \bibinfo {pages} {143} (\bibinfo {year} {1984})}\BibitemShut {NoStop}%
\bibitem [{\citenamefont {Brown}\ and\ \citenamefont
  {York}(1989)}]{Brown:1989ne}%
  \BibitemOpen
  \bibfield  {author} {\bibinfo {author} {\bibfnamefont {J.~D.}\ \bibnamefont
  {Brown}}\ and\ \bibinfo {author} {\bibfnamefont {J.~W.}\ \bibnamefont {York},
  \bibfnamefont {Jr.}},\ }\bibfield  {title} {\bibinfo {title} {{Jacobi's
  Action and the Recovery of Time in General Relativity}},\ }\href
  {https://doi.org/10.1103/PhysRevD.40.3312} {\bibfield  {journal} {\bibinfo
  {journal} {Phys. Rev. D}\ }\textbf {\bibinfo {volume} {40}},\ \bibinfo
  {pages} {3312} (\bibinfo {year} {1989})}\BibitemShut {NoStop}%
\bibitem [{\citenamefont {Ng}\ and\ \citenamefont {van Dam}(1990)}]{Ng:1990rw}%
  \BibitemOpen
  \bibfield  {author} {\bibinfo {author} {\bibfnamefont {Y.~J.}\ \bibnamefont
  {Ng}}\ and\ \bibinfo {author} {\bibfnamefont {H.}~\bibnamefont {van Dam}},\
  }\bibfield  {title} {\bibinfo {title} {{Possible solution to the cosmological
  constant problem}},\ }\href {https://doi.org/10.1103/PhysRevLett.65.1972}
  {\bibfield  {journal} {\bibinfo  {journal} {Phys. Rev. Lett.}\ }\textbf
  {\bibinfo {volume} {65}},\ \bibinfo {pages} {1972} (\bibinfo {year}
  {1990})}\BibitemShut {NoStop}%
\bibitem [{\citenamefont {Ng}\ and\ \citenamefont {van Dam}(1991)}]{Ng:1990xz}%
  \BibitemOpen
  \bibfield  {author} {\bibinfo {author} {\bibfnamefont {Y.~J.}\ \bibnamefont
  {Ng}}\ and\ \bibinfo {author} {\bibfnamefont {H.}~\bibnamefont {van Dam}},\
  }\bibfield  {title} {\bibinfo {title} {{Unimodular Theory of Gravity and the
  Cosmological Constant}},\ }\href {https://doi.org/10.1063/1.529283}
  {\bibfield  {journal} {\bibinfo  {journal} {J. Math. Phys.}\ }\textbf
  {\bibinfo {volume} {32}},\ \bibinfo {pages} {1337} (\bibinfo {year}
  {1991})}\BibitemShut {NoStop}%
\bibitem [{\citenamefont {Sorkin}(1997)}]{Sorkin:1997gi}%
  \BibitemOpen
  \bibfield  {author} {\bibinfo {author} {\bibfnamefont {R.~D.}\ \bibnamefont
  {Sorkin}},\ }\bibfield  {title} {\bibinfo {title} {{Forks in the road, on the
  way to quantum gravity}},\ }\href {https://doi.org/10.1007/BF02435709}
  {\bibfield  {journal} {\bibinfo  {journal} {Int. J. Theor. Phys.}\ }\textbf
  {\bibinfo {volume} {36}},\ \bibinfo {pages} {2759} (\bibinfo {year}
  {1997})},\ \Eprint {https://arxiv.org/abs/gr-qc/9706002}
  {arXiv:gr-qc/9706002} \BibitemShut {NoStop}%
\bibitem [{\citenamefont {Alvarez}\ and\ \citenamefont
  {Faedo}(2007)}]{Alvarez:2007nn}%
  \BibitemOpen
  \bibfield  {author} {\bibinfo {author} {\bibfnamefont {E.}~\bibnamefont
  {Alvarez}}\ and\ \bibinfo {author} {\bibfnamefont {A.~F.}\ \bibnamefont
  {Faedo}},\ }\bibfield  {title} {\bibinfo {title} {{Unimodular cosmology and
  the weight of energy}},\ }\href {https://doi.org/10.1103/PhysRevD.76.064013}
  {\bibfield  {journal} {\bibinfo  {journal} {Phys. Rev. D}\ }\textbf {\bibinfo
  {volume} {76}},\ \bibinfo {pages} {064013} (\bibinfo {year} {2007})},\
  \Eprint {https://arxiv.org/abs/hep-th/0702184} {arXiv:hep-th/0702184}
  \BibitemShut {NoStop}%
\bibitem [{\citenamefont {Smolin}(2009)}]{Smolin:2009ti}%
  \BibitemOpen
  \bibfield  {author} {\bibinfo {author} {\bibfnamefont {L.}~\bibnamefont
  {Smolin}},\ }\bibfield  {title} {\bibinfo {title} {{The Quantization of
  unimodular gravity and the cosmological constant problems}},\ }\href
  {https://doi.org/10.1103/PhysRevD.80.084003} {\bibfield  {journal} {\bibinfo
  {journal} {Phys. Rev. D}\ }\textbf {\bibinfo {volume} {80}},\ \bibinfo
  {pages} {084003} (\bibinfo {year} {2009})},\ \Eprint
  {https://arxiv.org/abs/0904.4841} {arXiv:0904.4841 [hep-th]} \BibitemShut
  {NoStop}%
\bibitem [{\citenamefont {Smolin}(2011)}]{Smolin:2010iq}%
  \BibitemOpen
  \bibfield  {author} {\bibinfo {author} {\bibfnamefont {L.}~\bibnamefont
  {Smolin}},\ }\bibfield  {title} {\bibinfo {title} {{Unimodular loop quantum
  gravity and the problems of time}},\ }\href
  {https://doi.org/10.1103/PhysRevD.84.044047} {\bibfield  {journal} {\bibinfo
  {journal} {Phys. Rev. D}\ }\textbf {\bibinfo {volume} {84}},\ \bibinfo
  {pages} {044047} (\bibinfo {year} {2011})},\ \Eprint
  {https://arxiv.org/abs/1008.1759} {arXiv:1008.1759 [hep-th]} \BibitemShut
  {NoStop}%
\bibitem [{\citenamefont {Alvarez}\ and\ \citenamefont
  {Vidal}(2010)}]{Alvarez:2010cg}%
  \BibitemOpen
  \bibfield  {author} {\bibinfo {author} {\bibfnamefont {E.}~\bibnamefont
  {Alvarez}}\ and\ \bibinfo {author} {\bibfnamefont {R.}~\bibnamefont
  {Vidal}},\ }\bibfield  {title} {\bibinfo {title} {{Weyl transverse gravity
  (WTDiff) and the cosmological constant}},\ }\href
  {https://doi.org/10.1103/PhysRevD.81.084057} {\bibfield  {journal} {\bibinfo
  {journal} {Phys. Rev. D}\ }\textbf {\bibinfo {volume} {81}},\ \bibinfo
  {pages} {084057} (\bibinfo {year} {2010})},\ \Eprint
  {https://arxiv.org/abs/1001.4458} {arXiv:1001.4458 [hep-th]} \BibitemShut
  {NoStop}%
\bibitem [{\citenamefont {Carballo-Rubio}(2015)}]{Carballo-Rubio:2015kaa}%
  \BibitemOpen
  \bibfield  {author} {\bibinfo {author} {\bibfnamefont {R.}~\bibnamefont
  {Carballo-Rubio}},\ }\bibfield  {title} {\bibinfo {title} {{Longitudinal
  diffeomorphisms obstruct the protection of vacuum energy}},\ }\href
  {https://doi.org/10.1103/PhysRevD.91.124071} {\bibfield  {journal} {\bibinfo
  {journal} {Phys. Rev. D}\ }\textbf {\bibinfo {volume} {91}},\ \bibinfo
  {pages} {124071} (\bibinfo {year} {2015})},\ \Eprint
  {https://arxiv.org/abs/1502.05278} {arXiv:1502.05278 [gr-qc]} \BibitemShut
  {NoStop}%
\bibitem [{\citenamefont {\'Alvarez}\ \emph
  {et~al.}(2015{\natexlab{a}})\citenamefont {\'Alvarez}, \citenamefont
  {Gonz\'alez-Mart\'\i{}n}, \citenamefont {Herrero-Valea},\ and\ \citenamefont
  {Mart\'\i{}n}}]{Alvarez:2015pla}%
  \BibitemOpen
  \bibfield  {author} {\bibinfo {author} {\bibfnamefont {E.}~\bibnamefont
  {\'Alvarez}}, \bibinfo {author} {\bibfnamefont {S.}~\bibnamefont
  {Gonz\'alez-Mart\'\i{}n}}, \bibinfo {author} {\bibfnamefont {M.}~\bibnamefont
  {Herrero-Valea}},\ and\ \bibinfo {author} {\bibfnamefont {C.~P.}\
  \bibnamefont {Mart\'\i{}n}},\ }\bibfield  {title} {\bibinfo {title}
  {{Unimodular Gravity Redux}},\ }\href
  {https://doi.org/10.1103/PhysRevD.92.061502} {\bibfield  {journal} {\bibinfo
  {journal} {Phys. Rev. D}\ }\textbf {\bibinfo {volume} {92}},\ \bibinfo
  {pages} {061502} (\bibinfo {year} {2015}{\natexlab{a}})},\ \Eprint
  {https://arxiv.org/abs/1505.00022} {arXiv:1505.00022 [hep-th]} \BibitemShut
  {NoStop}%
\bibitem [{\citenamefont {\'Alvarez}\ \emph
  {et~al.}(2015{\natexlab{b}})\citenamefont {\'Alvarez}, \citenamefont
  {Gonz\'alez-Mart\'\i{}n}, \citenamefont {Herrero-Valea},\ and\ \citenamefont
  {Mart\'\i{}n}}]{Alvarez:2015sba}%
  \BibitemOpen
  \bibfield  {author} {\bibinfo {author} {\bibfnamefont {E.}~\bibnamefont
  {\'Alvarez}}, \bibinfo {author} {\bibfnamefont {S.}~\bibnamefont
  {Gonz\'alez-Mart\'\i{}n}}, \bibinfo {author} {\bibfnamefont {M.}~\bibnamefont
  {Herrero-Valea}},\ and\ \bibinfo {author} {\bibfnamefont {C.~P.}\
  \bibnamefont {Mart\'\i{}n}},\ }\bibfield  {title} {\bibinfo {title} {{Quantum
  Corrections to Unimodular Gravity}},\ }\href
  {https://doi.org/10.1007/JHEP08(2015)078} {\bibfield  {journal} {\bibinfo
  {journal} {JHEP}\ }\textbf {\bibinfo {volume} {08}},\ \bibinfo {pages}
  {078}},\ \Eprint {https://arxiv.org/abs/1505.01995} {arXiv:1505.01995
  [hep-th]} \BibitemShut {NoStop}%
\bibitem [{\citenamefont {Percacci}(2018)}]{Percacci:2017fsy}%
  \BibitemOpen
  \bibfield  {author} {\bibinfo {author} {\bibfnamefont {R.}~\bibnamefont
  {Percacci}},\ }\bibfield  {title} {\bibinfo {title} {{Unimodular quantum
  gravity and the cosmological constant}},\ }\href
  {https://doi.org/10.1007/s10701-018-0189-5} {\bibfield  {journal} {\bibinfo
  {journal} {Found. Phys.}\ }\textbf {\bibinfo {volume} {48}},\ \bibinfo
  {pages} {1364} (\bibinfo {year} {2018})},\ \Eprint
  {https://arxiv.org/abs/1712.09903} {arXiv:1712.09903 [gr-qc]} \BibitemShut
  {NoStop}%
\bibitem [{\citenamefont {Fiol}\ and\ \citenamefont
  {Garriga}(2010)}]{Fiol:2008vk}%
  \BibitemOpen
  \bibfield  {author} {\bibinfo {author} {\bibfnamefont {B.}~\bibnamefont
  {Fiol}}\ and\ \bibinfo {author} {\bibfnamefont {J.}~\bibnamefont {Garriga}},\
  }\bibfield  {title} {\bibinfo {title} {{Semiclassical Unimodular Gravity}},\
  }\href {https://doi.org/10.1088/1475-7516/2010/08/015} {\bibfield  {journal}
  {\bibinfo  {journal} {JCAP}\ }\textbf {\bibinfo {volume} {08}},\ \bibinfo
  {pages} {015}},\ \Eprint {https://arxiv.org/abs/0809.1371} {arXiv:0809.1371
  [hep-th]} \BibitemShut {NoStop}%
\bibitem [{\citenamefont {Padilla}\ and\ \citenamefont
  {Saltas}(2015)}]{Padilla:2014yea}%
  \BibitemOpen
  \bibfield  {author} {\bibinfo {author} {\bibfnamefont {A.}~\bibnamefont
  {Padilla}}\ and\ \bibinfo {author} {\bibfnamefont {I.~D.}\ \bibnamefont
  {Saltas}},\ }\bibfield  {title} {\bibinfo {title} {{A note on classical and
  quantum unimodular gravity}},\ }\href
  {https://doi.org/10.1140/epjc/s10052-015-3767-0} {\bibfield  {journal}
  {\bibinfo  {journal} {Eur. Phys. J. C}\ }\textbf {\bibinfo {volume} {75}},\
  \bibinfo {pages} {561} (\bibinfo {year} {2015})},\ \Eprint
  {https://arxiv.org/abs/1409.3573} {arXiv:1409.3573 [gr-qc]} \BibitemShut
  {NoStop}%
\bibitem [{\citenamefont {Padilla}(2015)}]{Padilla:2015aaa}%
  \BibitemOpen
  \bibfield  {author} {\bibinfo {author} {\bibfnamefont {A.}~\bibnamefont
  {Padilla}},\ }\bibfield  {title} {\bibinfo {title} {{Lectures on the
  Cosmological Constant Problem}},\ }\href@noop {} {\  (\bibinfo {year}
  {2015})},\ \Eprint {https://arxiv.org/abs/1502.05296} {arXiv:1502.05296
  [hep-th]} \BibitemShut {NoStop}%
\bibitem [{\citenamefont {Hawking}(1978)}]{Hawking:1979zw}%
  \BibitemOpen
  \bibfield  {author} {\bibinfo {author} {\bibfnamefont {S.~W.}\ \bibnamefont
  {Hawking}},\ }\bibfield  {title} {\bibinfo {title} {{Space-Time Foam}},\
  }\href {https://doi.org/10.1016/0550-3213(78)90375-9} {\bibfield  {journal}
  {\bibinfo  {journal} {Nucl. Phys. B}\ }\textbf {\bibinfo {volume} {144}},\
  \bibinfo {pages} {349} (\bibinfo {year} {1978})}\BibitemShut {NoStop}%
\bibitem [{\citenamefont {Bufalo}\ \emph {et~al.}(2015)\citenamefont {Bufalo},
  \citenamefont {Oksanen},\ and\ \citenamefont {Tureanu}}]{Bufalo:2015wda}%
  \BibitemOpen
  \bibfield  {author} {\bibinfo {author} {\bibfnamefont {R.}~\bibnamefont
  {Bufalo}}, \bibinfo {author} {\bibfnamefont {M.}~\bibnamefont {Oksanen}},\
  and\ \bibinfo {author} {\bibfnamefont {A.}~\bibnamefont {Tureanu}},\
  }\bibfield  {title} {\bibinfo {title} {{How unimodular gravity theories
  differ from general relativity at quantum level}},\ }\href
  {https://doi.org/10.1140/epjc/s10052-015-3683-3} {\bibfield  {journal}
  {\bibinfo  {journal} {Eur. Phys. J. C}\ }\textbf {\bibinfo {volume} {75}},\
  \bibinfo {pages} {477} (\bibinfo {year} {2015})},\ \Eprint
  {https://arxiv.org/abs/1505.04978} {arXiv:1505.04978 [hep-th]} \BibitemShut
  {NoStop}%
\bibitem [{\citenamefont {Upadhyay}\ \emph {et~al.}(2017)\citenamefont
  {Upadhyay}, \citenamefont {Oksanen},\ and\ \citenamefont
  {Bufalo}}]{Upadhyay:2015fna}%
  \BibitemOpen
  \bibfield  {author} {\bibinfo {author} {\bibfnamefont {S.}~\bibnamefont
  {Upadhyay}}, \bibinfo {author} {\bibfnamefont {M.}~\bibnamefont {Oksanen}},\
  and\ \bibinfo {author} {\bibfnamefont {R.}~\bibnamefont {Bufalo}},\
  }\bibfield  {title} {\bibinfo {title} {{BRST Quantization of Unimodular
  Gravity}},\ }\href {https://doi.org/10.1007/s13538-017-0500-5} {\bibfield
  {journal} {\bibinfo  {journal} {Braz. J. Phys.}\ }\textbf {\bibinfo {volume}
  {47}},\ \bibinfo {pages} {350} (\bibinfo {year} {2017})},\ \Eprint
  {https://arxiv.org/abs/1510.00188} {arXiv:1510.00188 [hep-th]} \BibitemShut
  {NoStop}%
\bibitem [{\citenamefont {Eichhorn}(2013)}]{Eichhorn:2013xr}%
  \BibitemOpen
  \bibfield  {author} {\bibinfo {author} {\bibfnamefont {A.}~\bibnamefont
  {Eichhorn}},\ }\bibfield  {title} {\bibinfo {title} {{On unimodular quantum
  gravity}},\ }\href {https://doi.org/10.1088/0264-9381/30/11/115016}
  {\bibfield  {journal} {\bibinfo  {journal} {Class. Quant. Grav.}\ }\textbf
  {\bibinfo {volume} {30}},\ \bibinfo {pages} {115016} (\bibinfo {year}
  {2013})},\ \Eprint {https://arxiv.org/abs/1301.0879} {arXiv:1301.0879
  [gr-qc]} \BibitemShut {NoStop}%
\bibitem [{\citenamefont {Eichhorn}(2015)}]{Eichhorn:2015bna}%
  \BibitemOpen
  \bibfield  {author} {\bibinfo {author} {\bibfnamefont {A.}~\bibnamefont
  {Eichhorn}},\ }\bibfield  {title} {\bibinfo {title} {{The Renormalization
  Group flow of unimodular f(R) gravity}},\ }\href
  {https://doi.org/10.1007/JHEP04(2015)096} {\bibfield  {journal} {\bibinfo
  {journal} {JHEP}\ }\textbf {\bibinfo {volume} {04}},\ \bibinfo {pages}
  {096}},\ \Eprint {https://arxiv.org/abs/1501.05848} {arXiv:1501.05848
  [gr-qc]} \BibitemShut {NoStop}%
\bibitem [{\citenamefont {Saltas}(2014)}]{Saltas:2014cta}%
  \BibitemOpen
  \bibfield  {author} {\bibinfo {author} {\bibfnamefont {I.~D.}\ \bibnamefont
  {Saltas}},\ }\bibfield  {title} {\bibinfo {title} {{UV structure of quantum
  unimodular gravity}},\ }\href {https://doi.org/10.1103/PhysRevD.90.124052}
  {\bibfield  {journal} {\bibinfo  {journal} {Phys. Rev. D}\ }\textbf {\bibinfo
  {volume} {90}},\ \bibinfo {pages} {124052} (\bibinfo {year} {2014})},\
  \Eprint {https://arxiv.org/abs/1410.6163} {arXiv:1410.6163 [hep-th]}
  \BibitemShut {NoStop}%
\bibitem [{\citenamefont {Benedetti}(2016)}]{Benedetti:2015zsw}%
  \BibitemOpen
  \bibfield  {author} {\bibinfo {author} {\bibfnamefont {D.}~\bibnamefont
  {Benedetti}},\ }\bibfield  {title} {\bibinfo {title} {{Essential nature of
  Newton\textquoteright{}s constant in unimodular gravity}},\ }\href
  {https://doi.org/10.1007/s10714-016-2060-3} {\bibfield  {journal} {\bibinfo
  {journal} {Gen. Rel. Grav.}\ }\textbf {\bibinfo {volume} {48}},\ \bibinfo
  {pages} {68} (\bibinfo {year} {2016})},\ \Eprint
  {https://arxiv.org/abs/1511.06560} {arXiv:1511.06560 [hep-th]} \BibitemShut
  {NoStop}%
\bibitem [{\citenamefont {Burger}\ \emph {et~al.}(2015)\citenamefont {Burger},
  \citenamefont {Ellis}, \citenamefont {Murugan},\ and\ \citenamefont
  {Weltman}}]{Burger:2015kie}%
  \BibitemOpen
  \bibfield  {author} {\bibinfo {author} {\bibfnamefont {D.~J.}\ \bibnamefont
  {Burger}}, \bibinfo {author} {\bibfnamefont {G.~F.~R.}\ \bibnamefont
  {Ellis}}, \bibinfo {author} {\bibfnamefont {J.}~\bibnamefont {Murugan}},\
  and\ \bibinfo {author} {\bibfnamefont {A.}~\bibnamefont {Weltman}},\
  }\bibfield  {title} {\bibinfo {title} {{The KLT relations in unimodular
  gravity}},\ }\href@noop {} {\  (\bibinfo {year} {2015})},\ \Eprint
  {https://arxiv.org/abs/1511.08517} {arXiv:1511.08517 [hep-th]} \BibitemShut
  {NoStop}%
\bibitem [{\citenamefont {Alvarez}\ \emph {et~al.}(2016)\citenamefont
  {Alvarez}, \citenamefont {Gonzalez-Martin},\ and\ \citenamefont
  {Martin}}]{Alvarez:2016uog}%
  \BibitemOpen
  \bibfield  {author} {\bibinfo {author} {\bibfnamefont {E.}~\bibnamefont
  {Alvarez}}, \bibinfo {author} {\bibfnamefont {S.}~\bibnamefont
  {Gonzalez-Martin}},\ and\ \bibinfo {author} {\bibfnamefont {C.~P.}\
  \bibnamefont {Martin}},\ }\bibfield  {title} {\bibinfo {title} {{Unimodular
  Trees versus Einstein Trees}},\ }\href
  {https://doi.org/10.1140/epjc/s10052-016-4384-2} {\bibfield  {journal}
  {\bibinfo  {journal} {Eur. Phys. J. C}\ }\textbf {\bibinfo {volume} {76}},\
  \bibinfo {pages} {554} (\bibinfo {year} {2016})},\ \Eprint
  {https://arxiv.org/abs/1605.02667} {arXiv:1605.02667 [hep-th]} \BibitemShut
  {NoStop}%
\bibitem [{\citenamefont {de~León~Ardón}\ \emph {et~al.}(2018)\citenamefont
  {de~León~Ardón}, \citenamefont {Ohta},\ and\ \citenamefont
  {Percacci}}]{Ardon:2017atk}%
  \BibitemOpen
  \bibfield  {author} {\bibinfo {author} {\bibfnamefont {R.}~\bibnamefont
  {de~León~Ardón}}, \bibinfo {author} {\bibfnamefont {N.}~\bibnamefont
  {Ohta}},\ and\ \bibinfo {author} {\bibfnamefont {R.}~\bibnamefont
  {Percacci}},\ }\bibfield  {title} {\bibinfo {title} {{Path integral of
  unimodular gravity}},\ }\href {https://doi.org/10.1103/PhysRevD.97.026007}
  {\bibfield  {journal} {\bibinfo  {journal} {Phys. Rev. D}\ }\textbf {\bibinfo
  {volume} {97}},\ \bibinfo {pages} {026007} (\bibinfo {year} {2018})},\
  \Eprint {https://arxiv.org/abs/1710.02457} {arXiv:1710.02457 [gr-qc]}
  \BibitemShut {NoStop}%
\bibitem [{\citenamefont {Gonz\'alez-Mart\'\i{}n}\ and\ \citenamefont
  {Martin}(2018)}]{Gonzalez-Martin:2017fwz}%
  \BibitemOpen
  \bibfield  {author} {\bibinfo {author} {\bibfnamefont {S.}~\bibnamefont
  {Gonz\'alez-Mart\'\i{}n}}\ and\ \bibinfo {author} {\bibfnamefont {C.~P.}\
  \bibnamefont {Martin}},\ }\bibfield  {title} {\bibinfo {title} {{Unimodular
  Gravity and General Relativity UV divergent contributions to the scattering
  of massive scalar particles}},\ }\href
  {https://doi.org/10.1088/1475-7516/2018/01/028} {\bibfield  {journal}
  {\bibinfo  {journal} {JCAP}\ }\textbf {\bibinfo {volume} {01}},\ \bibinfo
  {pages} {028}},\ \Eprint {https://arxiv.org/abs/1711.08009} {arXiv:1711.08009
  [hep-th]} \BibitemShut {NoStop}%
\bibitem [{\citenamefont {Gonzalez-Martin}\ and\ \citenamefont
  {Martin}(2018)}]{Gonzalez-Martin:2018dmy}%
  \BibitemOpen
  \bibfield  {author} {\bibinfo {author} {\bibfnamefont {S.}~\bibnamefont
  {Gonzalez-Martin}}\ and\ \bibinfo {author} {\bibfnamefont {C.~P.}\
  \bibnamefont {Martin}},\ }\bibfield  {title} {\bibinfo {title} {{Scattering
  of fermions in the Yukawa theory coupled to Unimodular Gravity}},\ }\href
  {https://doi.org/10.1140/epjc/s10052-018-5734-z} {\bibfield  {journal}
  {\bibinfo  {journal} {Eur. Phys. J. C}\ }\textbf {\bibinfo {volume} {78}},\
  \bibinfo {pages} {236} (\bibinfo {year} {2018})},\ \Eprint
  {https://arxiv.org/abs/1802.03755} {arXiv:1802.03755 [hep-th]} \BibitemShut
  {NoStop}%
\bibitem [{\citenamefont {De~Brito}\ \emph {et~al.}(2019)\citenamefont
  {De~Brito}, \citenamefont {Eichhorn},\ and\ \citenamefont
  {Pereira}}]{deBrito:2019umw}%
  \BibitemOpen
  \bibfield  {author} {\bibinfo {author} {\bibfnamefont {G.~P.}\ \bibnamefont
  {De~Brito}}, \bibinfo {author} {\bibfnamefont {A.}~\bibnamefont {Eichhorn}},\
  and\ \bibinfo {author} {\bibfnamefont {A.~D.}\ \bibnamefont {Pereira}},\
  }\bibfield  {title} {\bibinfo {title} {{A link that matters: Towards
  phenomenological tests of unimodular asymptotic safety}},\ }\href
  {https://doi.org/10.1007/JHEP09(2019)100} {\bibfield  {journal} {\bibinfo
  {journal} {JHEP}\ }\textbf {\bibinfo {volume} {09}},\ \bibinfo {pages}
  {100}},\ \Eprint {https://arxiv.org/abs/1907.11173} {arXiv:1907.11173
  [hep-th]} \BibitemShut {NoStop}%
\bibitem [{\citenamefont {Yamashita}(2020)}]{Yamashita:2020ixd}%
  \BibitemOpen
  \bibfield  {author} {\bibinfo {author} {\bibfnamefont {S.}~\bibnamefont
  {Yamashita}},\ }\bibfield  {title} {\bibinfo {title} {{Hamiltonian analysis
  of unimodular gravity and its quantization in the connection
  representation}},\ }\href {https://doi.org/10.1103/PhysRevD.101.086007}
  {\bibfield  {journal} {\bibinfo  {journal} {Phys. Rev. D}\ }\textbf {\bibinfo
  {volume} {101}},\ \bibinfo {pages} {086007} (\bibinfo {year} {2020})},\
  \Eprint {https://arxiv.org/abs/2003.05083} {arXiv:2003.05083 [gr-qc]}
  \BibitemShut {NoStop}%
\bibitem [{\citenamefont {Baulieu}(2020{\natexlab{a}})}]{Baulieu:2020obv}%
  \BibitemOpen
  \bibfield  {author} {\bibinfo {author} {\bibfnamefont {L.}~\bibnamefont
  {Baulieu}},\ }\bibfield  {title} {\bibinfo {title} {{Unimodular Gauge in
  Perturbative Gravity and Supergravity}},\ }\href
  {https://doi.org/10.1016/j.physletb.2020.135591} {\bibfield  {journal}
  {\bibinfo  {journal} {Phys. Lett. B}\ }\textbf {\bibinfo {volume} {808}},\
  \bibinfo {pages} {135591} (\bibinfo {year} {2020}{\natexlab{a}})},\ \Eprint
  {https://arxiv.org/abs/2004.05950} {arXiv:2004.05950 [hep-th]} \BibitemShut
  {NoStop}%
\bibitem [{\citenamefont {Baulieu}(2020{\natexlab{b}})}]{Baulieu:2020rpv}%
  \BibitemOpen
  \bibfield  {author} {\bibinfo {author} {\bibfnamefont {L.}~\bibnamefont
  {Baulieu}},\ }\bibfield  {title} {\bibinfo {title} {{Unimodular Gauge and ADM
  Gravity Path Integral}},\ }\href@noop {} {\  (\bibinfo {year}
  {2020}{\natexlab{b}})},\ \Eprint {https://arxiv.org/abs/2012.01116}
  {arXiv:2012.01116 [hep-th]} \BibitemShut {NoStop}%
\bibitem [{\citenamefont {Herrero-Valea}\ and\ \citenamefont
  {Santos-Garcia}(2020)}]{Herrero-Valea:2020xaq}%
  \BibitemOpen
  \bibfield  {author} {\bibinfo {author} {\bibfnamefont {M.}~\bibnamefont
  {Herrero-Valea}}\ and\ \bibinfo {author} {\bibfnamefont {R.}~\bibnamefont
  {Santos-Garcia}},\ }\bibfield  {title} {\bibinfo {title} {{Non-minimal Tinges
  of Unimodular Gravity}},\ }\href {https://doi.org/10.1007/JHEP09(2020)041}
  {\bibfield  {journal} {\bibinfo  {journal} {JHEP}\ }\textbf {\bibinfo
  {volume} {09}},\ \bibinfo {pages} {041}},\ \Eprint
  {https://arxiv.org/abs/2006.06698} {arXiv:2006.06698 [hep-th]} \BibitemShut
  {NoStop}%
\bibitem [{\citenamefont {de~Brito}\ and\ \citenamefont
  {Pereira}(2020)}]{deBrito:2020rwu}%
  \BibitemOpen
  \bibfield  {author} {\bibinfo {author} {\bibfnamefont {G.~P.}\ \bibnamefont
  {de~Brito}}\ and\ \bibinfo {author} {\bibfnamefont {A.~D.}\ \bibnamefont
  {Pereira}},\ }\bibfield  {title} {\bibinfo {title} {{Unimodular quantum
  gravity: Steps beyond perturbation theory}},\ }\href
  {https://doi.org/10.1007/JHEP09(2020)196} {\bibfield  {journal} {\bibinfo
  {journal} {JHEP}\ }\textbf {\bibinfo {volume} {09}},\ \bibinfo {pages}
  {196}},\ \Eprint {https://arxiv.org/abs/2007.05589} {arXiv:2007.05589
  [hep-th]} \BibitemShut {NoStop}%
\bibitem [{\citenamefont {de~Brito}\ \emph {et~al.}(2020)\citenamefont
  {de~Brito}, \citenamefont {Pereira},\ and\ \citenamefont
  {Vieira}}]{deBrito:2020xhy}%
  \BibitemOpen
  \bibfield  {author} {\bibinfo {author} {\bibfnamefont {G.~P.}\ \bibnamefont
  {de~Brito}}, \bibinfo {author} {\bibfnamefont {A.~D.}\ \bibnamefont
  {Pereira}},\ and\ \bibinfo {author} {\bibfnamefont {A.~F.}\ \bibnamefont
  {Vieira}},\ }\bibfield  {title} {\bibinfo {title} {{Exploring new corners of
  asymptotically safe unimodular quantum gravity}},\ }\href@noop {} {\
  (\bibinfo {year} {2020})},\ \Eprint {https://arxiv.org/abs/2012.08904}
  {arXiv:2012.08904 [hep-th]} \BibitemShut {NoStop}%
\bibitem [{\citenamefont {Ohta}\ \emph {et~al.}(2018)\citenamefont {Ohta},
  \citenamefont {Percacci},\ and\ \citenamefont {Pereira}}]{Ohta:2018sze}%
  \BibitemOpen
  \bibfield  {author} {\bibinfo {author} {\bibfnamefont {N.}~\bibnamefont
  {Ohta}}, \bibinfo {author} {\bibfnamefont {R.}~\bibnamefont {Percacci}},\
  and\ \bibinfo {author} {\bibfnamefont {A.~D.}\ \bibnamefont {Pereira}},\
  }\bibfield  {title} {\bibinfo {title} {{$f(R, R_{\mu\nu}^2)$ at one loop}},\
  }\href {https://doi.org/10.1103/PhysRevD.97.104039} {\bibfield  {journal}
  {\bibinfo  {journal} {Phys. Rev. D}\ }\textbf {\bibinfo {volume} {97}},\
  \bibinfo {pages} {104039} (\bibinfo {year} {2018})},\ \Eprint
  {https://arxiv.org/abs/1804.01608} {arXiv:1804.01608 [hep-th]} \BibitemShut
  {NoStop}%
\bibitem [{\citenamefont {Gonzalez-Martin}\ and\ \citenamefont
  {Martin}(2017)}]{Gonzalez-Martin:2017bvw}%
  \BibitemOpen
  \bibfield  {author} {\bibinfo {author} {\bibfnamefont {S.}~\bibnamefont
  {Gonzalez-Martin}}\ and\ \bibinfo {author} {\bibfnamefont {C.~P.}\
  \bibnamefont {Martin}},\ }\bibfield  {title} {\bibinfo {title} {{Do the
  gravitational corrections to the beta functions of the quartic and Yukawa
  couplings have an intrinsic physical meaning?}},\ }\href
  {https://doi.org/10.1016/j.physletb.2017.09.011} {\bibfield  {journal}
  {\bibinfo  {journal} {Phys. Lett. B}\ }\textbf {\bibinfo {volume} {773}},\
  \bibinfo {pages} {585} (\bibinfo {year} {2017})},\ \Eprint
  {https://arxiv.org/abs/1707.06667} {arXiv:1707.06667 [hep-th]} \BibitemShut
  {NoStop}%
\bibitem [{\citenamefont {Ferrari}(2014)}]{Ferrari:2013aza}%
  \BibitemOpen
  \bibfield  {author} {\bibinfo {author} {\bibfnamefont {F.}~\bibnamefont
  {Ferrari}},\ }\bibfield  {title} {\bibinfo {title} {{Partial Gauge Fixing and
  Equivariant Cohomology}},\ }\href
  {https://doi.org/10.1103/PhysRevD.89.105018} {\bibfield  {journal} {\bibinfo
  {journal} {Phys. Rev. D}\ }\textbf {\bibinfo {volume} {89}},\ \bibinfo
  {pages} {105018} (\bibinfo {year} {2014})},\ \Eprint
  {https://arxiv.org/abs/1308.6802} {arXiv:1308.6802 [hep-th]} \BibitemShut
  {NoStop}%
\bibitem [{\citenamefont {Ohta}\ \emph {et~al.}(2016)\citenamefont {Ohta},
  \citenamefont {Percacci},\ and\ \citenamefont {Pereira}}]{Ohta:2016npm}%
  \BibitemOpen
  \bibfield  {author} {\bibinfo {author} {\bibfnamefont {N.}~\bibnamefont
  {Ohta}}, \bibinfo {author} {\bibfnamefont {R.}~\bibnamefont {Percacci}},\
  and\ \bibinfo {author} {\bibfnamefont {A.}~\bibnamefont {Pereira}},\
  }\bibfield  {title} {\bibinfo {title} {{Gauges and functional measures in
  quantum gravity I: Einstein theory}},\ }\href
  {https://doi.org/10.1007/JHEP06(2016)115} {\bibfield  {journal} {\bibinfo
  {journal} {JHEP}\ }\textbf {\bibinfo {volume} {06}},\ \bibinfo {pages}
  {115}},\ \Eprint {https://arxiv.org/abs/1605.00454} {arXiv:1605.00454
  [hep-th]} \BibitemShut {NoStop}%
\bibitem [{\citenamefont {Alvarez}\ \emph {et~al.}(2008)\citenamefont
  {Alvarez}, \citenamefont {Faedo},\ and\ \citenamefont
  {Lopez-Villarejo}}]{Alvarez:2008zw}%
  \BibitemOpen
  \bibfield  {author} {\bibinfo {author} {\bibfnamefont {E.}~\bibnamefont
  {Alvarez}}, \bibinfo {author} {\bibfnamefont {A.~F.}\ \bibnamefont {Faedo}},\
  and\ \bibinfo {author} {\bibfnamefont {J.~J.}\ \bibnamefont
  {Lopez-Villarejo}},\ }\bibfield  {title} {\bibinfo {title} {{Ultraviolet
  behavior of transverse gravity}},\ }\href
  {https://doi.org/10.1088/1126-6708/2008/10/023} {\bibfield  {journal}
  {\bibinfo  {journal} {JHEP}\ }\textbf {\bibinfo {volume} {10}},\ \bibinfo
  {pages} {023}},\ \Eprint {https://arxiv.org/abs/0807.1293} {arXiv:0807.1293
  [hep-th]} \BibitemShut {NoStop}%
\bibitem [{\citenamefont {Narain}\ and\ \citenamefont
  {Percacci}(2010)}]{Narain:2009fy}%
  \BibitemOpen
  \bibfield  {author} {\bibinfo {author} {\bibfnamefont {G.}~\bibnamefont
  {Narain}}\ and\ \bibinfo {author} {\bibfnamefont {R.}~\bibnamefont
  {Percacci}},\ }\bibfield  {title} {\bibinfo {title} {{Renormalization Group
  Flow in Scalar-Tensor Theories. I}},\ }\href
  {https://doi.org/10.1088/0264-9381/27/7/075001} {\bibfield  {journal}
  {\bibinfo  {journal} {Class. Quant. Grav.}\ }\textbf {\bibinfo {volume}
  {27}},\ \bibinfo {pages} {075001} (\bibinfo {year} {2010})},\ \Eprint
  {https://arxiv.org/abs/0911.0386} {arXiv:0911.0386 [hep-th]} \BibitemShut
  {NoStop}%
\bibitem [{\citenamefont {Percacci}\ and\ \citenamefont
  {Vacca}(2015)}]{Percacci:2015wwa}%
  \BibitemOpen
  \bibfield  {author} {\bibinfo {author} {\bibfnamefont {R.}~\bibnamefont
  {Percacci}}\ and\ \bibinfo {author} {\bibfnamefont {G.~P.}\ \bibnamefont
  {Vacca}},\ }\bibfield  {title} {\bibinfo {title} {{Search of scaling
  solutions in scalar-tensor gravity}},\ }\href
  {https://doi.org/10.1140/epjc/s10052-015-3410-0} {\bibfield  {journal}
  {\bibinfo  {journal} {Eur. Phys. J. C}\ }\textbf {\bibinfo {volume} {75}},\
  \bibinfo {pages} {188} (\bibinfo {year} {2015})},\ \Eprint
  {https://arxiv.org/abs/1501.00888} {arXiv:1501.00888 [hep-th]} \BibitemShut
  {NoStop}%
\bibitem [{\citenamefont {Fr\"ob}(2018)}]{Frob:2017lnt}%
  \BibitemOpen
  \bibfield  {author} {\bibinfo {author} {\bibfnamefont {M.~B.}\ \bibnamefont
  {Fr\"ob}},\ }\bibfield  {title} {\bibinfo {title} {{Gauge-invariant quantum
  gravitational corrections to correlation functions}},\ }\href
  {https://doi.org/10.1088/1361-6382/aaa74c} {\bibfield  {journal} {\bibinfo
  {journal} {Class. Quant. Grav.}\ }\textbf {\bibinfo {volume} {35}},\ \bibinfo
  {pages} {055006} (\bibinfo {year} {2018})},\ \Eprint
  {https://arxiv.org/abs/1710.00839} {arXiv:1710.00839 [gr-qc]} \BibitemShut
  {NoStop}%
\bibitem [{\citenamefont {Steinwachs}\ and\ \citenamefont
  {Kamenshchik}(2011)}]{Steinwachs:2011zs}%
  \BibitemOpen
  \bibfield  {author} {\bibinfo {author} {\bibfnamefont {C.~F.}\ \bibnamefont
  {Steinwachs}}\ and\ \bibinfo {author} {\bibfnamefont {A.~Y.}\ \bibnamefont
  {Kamenshchik}},\ }\bibfield  {title} {\bibinfo {title} {{One-loop divergences
  for gravity non-minimally coupled to a multiplet of scalar fields:
  calculation in the Jordan frame. I. The main results}},\ }\href
  {https://doi.org/10.1103/PhysRevD.84.024026} {\bibfield  {journal} {\bibinfo
  {journal} {Phys. Rev. D}\ }\textbf {\bibinfo {volume} {84}},\ \bibinfo
  {pages} {024026} (\bibinfo {year} {2011})},\ \Eprint
  {https://arxiv.org/abs/1101.5047} {arXiv:1101.5047 [gr-qc]} \BibitemShut
  {NoStop}%
\bibitem [{\citenamefont {Shapiro}\ and\ \citenamefont
  {Takata}(1995)}]{Shapiro:1995yc}%
  \BibitemOpen
  \bibfield  {author} {\bibinfo {author} {\bibfnamefont {I.~L.}\ \bibnamefont
  {Shapiro}}\ and\ \bibinfo {author} {\bibfnamefont {H.}~\bibnamefont
  {Takata}},\ }\bibfield  {title} {\bibinfo {title} {{One loop renormalization
  of the four-dimensional theory for quantum dilaton gravity}},\ }\href
  {https://doi.org/10.1103/PhysRevD.52.2162} {\bibfield  {journal} {\bibinfo
  {journal} {Phys. Rev. D}\ }\textbf {\bibinfo {volume} {52}},\ \bibinfo
  {pages} {2162} (\bibinfo {year} {1995})},\ \Eprint
  {https://arxiv.org/abs/hep-th/9502111} {arXiv:hep-th/9502111} \BibitemShut
  {NoStop}%
\bibitem [{\citenamefont {Pagani}\ and\ \citenamefont
  {Reuter}(2019)}]{Pagani:2019vfm}%
  \BibitemOpen
  \bibfield  {author} {\bibinfo {author} {\bibfnamefont {C.}~\bibnamefont
  {Pagani}}\ and\ \bibinfo {author} {\bibfnamefont {M.}~\bibnamefont
  {Reuter}},\ }\bibfield  {title} {\bibinfo {title} {{Background Independent
  Quantum Field Theory and Gravitating Vacuum Fluctuations}},\ }\href
  {https://doi.org/10.1016/j.aop.2019.167972} {\bibfield  {journal} {\bibinfo
  {journal} {Annals Phys.}\ }\textbf {\bibinfo {volume} {411}},\ \bibinfo
  {pages} {167972} (\bibinfo {year} {2019})},\ \Eprint
  {https://arxiv.org/abs/1906.02507} {arXiv:1906.02507 [gr-qc]} \BibitemShut
  {NoStop}%
\bibitem [{\citenamefont {Pagani}\ and\ \citenamefont
  {Reuter}(2020)}]{Pagani:2020say}%
  \BibitemOpen
  \bibfield  {author} {\bibinfo {author} {\bibfnamefont {C.}~\bibnamefont
  {Pagani}}\ and\ \bibinfo {author} {\bibfnamefont {M.}~\bibnamefont
  {Reuter}},\ }\bibfield  {title} {\bibinfo {title} {{Why the Cosmological
  Constant Seems to Hardly Care About Quantum Vacuum Fluctuations: Surprises
  From Background Independent Coarse Graining}},\ }\href
  {https://doi.org/10.3389/fphy.2020.00214} {\bibfield  {journal} {\bibinfo
  {journal} {Front. in Phys.}\ }\textbf {\bibinfo {volume} {8}},\ \bibinfo
  {pages} {214} (\bibinfo {year} {2020})}\BibitemShut {NoStop}%
\bibitem [{\citenamefont {Becker}\ and\ \citenamefont
  {Reuter}(2020)}]{Becker:2020mjl}%
  \BibitemOpen
  \bibfield  {author} {\bibinfo {author} {\bibfnamefont {M.}~\bibnamefont
  {Becker}}\ and\ \bibinfo {author} {\bibfnamefont {M.}~\bibnamefont
  {Reuter}},\ }\bibfield  {title} {\bibinfo {title} {{Background Independent
  Field Quantization with Sequences of Gravity-Coupled Approximants}},\ }\href
  {https://doi.org/10.1103/PhysRevD.102.125001} {\bibfield  {journal} {\bibinfo
   {journal} {Phys. Rev. D}\ }\textbf {\bibinfo {volume} {102}},\ \bibinfo
  {pages} {125001} (\bibinfo {year} {2020})},\ \Eprint
  {https://arxiv.org/abs/2008.09430} {arXiv:2008.09430 [gr-qc]} \BibitemShut
  {NoStop}%
\bibitem [{\citenamefont {Donoghue}(2020)}]{Donoghue:2020hoh}%
  \BibitemOpen
  \bibfield  {author} {\bibinfo {author} {\bibfnamefont {J.~F.}\ \bibnamefont
  {Donoghue}},\ }\bibfield  {title} {\bibinfo {title} {{The cosmological
  constant and the use of cutoffs}},\ }\href@noop {} {\  (\bibinfo {year}
  {2020})},\ \Eprint {https://arxiv.org/abs/2009.00728} {arXiv:2009.00728
  [hep-th]} \BibitemShut {NoStop}%
\bibitem [{\citenamefont {Percacci}(2017)}]{Percacci:2017fkn}%
  \BibitemOpen
  \bibfield  {author} {\bibinfo {author} {\bibfnamefont {R.}~\bibnamefont
  {Percacci}},\ }\href {https://doi.org/10.1142/10369} {\emph {\bibinfo {title}
  {{An Introduction to Covariant Quantum Gravity and Asymptotic Safety}}}},\
  \bibinfo {series} {100 Years of General Relativity}, Vol.~\bibinfo {volume}
  {3}\ (\bibinfo  {publisher} {World Scientific},\ \bibinfo {year}
  {2017})\BibitemShut {NoStop}%
\bibitem [{\citenamefont {Dupuis}\ \emph {et~al.}(2020)\citenamefont {Dupuis},
  \citenamefont {Canet}, \citenamefont {Eichhorn}, \citenamefont {Metzner},
  \citenamefont {Pawlowski}, \citenamefont {Tissier},\ and\ \citenamefont
  {Wschebor}}]{Dupuis:2020fhh}%
  \BibitemOpen
  \bibfield  {author} {\bibinfo {author} {\bibfnamefont {N.}~\bibnamefont
  {Dupuis}}, \bibinfo {author} {\bibfnamefont {L.}~\bibnamefont {Canet}},
  \bibinfo {author} {\bibfnamefont {A.}~\bibnamefont {Eichhorn}}, \bibinfo
  {author} {\bibfnamefont {W.}~\bibnamefont {Metzner}}, \bibinfo {author}
  {\bibfnamefont {J.~M.}\ \bibnamefont {Pawlowski}}, \bibinfo {author}
  {\bibfnamefont {M.}~\bibnamefont {Tissier}},\ and\ \bibinfo {author}
  {\bibfnamefont {N.}~\bibnamefont {Wschebor}},\ }\bibfield  {title} {\bibinfo
  {title} {{The nonperturbative functional renormalization group and its
  applications}}\ }\href {https://doi.org/10.1016/j.physrep.2021.01.001}
  {10.1016/j.physrep.2021.01.001} (\bibinfo {year} {2020}),\ \Eprint
  {https://arxiv.org/abs/2006.04853} {arXiv:2006.04853 [cond-mat.stat-mech]}
  \BibitemShut {NoStop}%
\bibitem [{\citenamefont {Pawlowski}(2007)}]{Pawlowski:2005xe}%
  \BibitemOpen
  \bibfield  {author} {\bibinfo {author} {\bibfnamefont {J.~M.}\ \bibnamefont
  {Pawlowski}},\ }\bibfield  {title} {\bibinfo {title} {{Aspects of the
  functional renormalisation group}},\ }\href
  {https://doi.org/10.1016/j.aop.2007.01.007} {\bibfield  {journal} {\bibinfo
  {journal} {Annals Phys.}\ }\textbf {\bibinfo {volume} {322}},\ \bibinfo
  {pages} {2831} (\bibinfo {year} {2007})},\ \Eprint
  {https://arxiv.org/abs/hep-th/0512261} {arXiv:hep-th/0512261} \BibitemShut
  {NoStop}%
\bibitem [{\citenamefont {Gies}(2012)}]{Gies:2006wv}%
  \BibitemOpen
  \bibfield  {author} {\bibinfo {author} {\bibfnamefont {H.}~\bibnamefont
  {Gies}},\ }\bibfield  {title} {\bibinfo {title} {{Introduction to the
  functional RG and applications to gauge theories}},\ }\href
  {https://doi.org/10.1007/978-3-642-27320-9_6} {\bibfield  {journal} {\bibinfo
   {journal} {Lect. Notes Phys.}\ }\textbf {\bibinfo {volume} {852}},\ \bibinfo
  {pages} {287} (\bibinfo {year} {2012})},\ \Eprint
  {https://arxiv.org/abs/hep-ph/0611146} {arXiv:hep-ph/0611146} \BibitemShut
  {NoStop}%
\bibitem [{\citenamefont {Baldazzi}\ \emph {et~al.}(2020)\citenamefont
  {Baldazzi}, \citenamefont {Percacci},\ and\ \citenamefont
  {Zambelli}}]{Baldazzi:2020vxk}%
  \BibitemOpen
  \bibfield  {author} {\bibinfo {author} {\bibfnamefont {A.}~\bibnamefont
  {Baldazzi}}, \bibinfo {author} {\bibfnamefont {R.}~\bibnamefont {Percacci}},\
  and\ \bibinfo {author} {\bibfnamefont {L.}~\bibnamefont {Zambelli}},\
  }\bibfield  {title} {\bibinfo {title} {{Functional Renormalization and
  $\overline{\text{MS}}$}},\ }\href@noop {} {\  (\bibinfo {year} {2020})},\
  \Eprint {https://arxiv.org/abs/2009.03255} {arXiv:2009.03255 [hep-th]}
  \BibitemShut {NoStop}%
\bibitem [{\citenamefont {Codello}\ \emph {et~al.}(2009)\citenamefont
  {Codello}, \citenamefont {Percacci},\ and\ \citenamefont
  {Rahmede}}]{Codello:2008vh}%
  \BibitemOpen
  \bibfield  {author} {\bibinfo {author} {\bibfnamefont {A.}~\bibnamefont
  {Codello}}, \bibinfo {author} {\bibfnamefont {R.}~\bibnamefont {Percacci}},\
  and\ \bibinfo {author} {\bibfnamefont {C.}~\bibnamefont {Rahmede}},\
  }\bibfield  {title} {\bibinfo {title} {{Investigating the Ultraviolet
  Properties of Gravity with a Wilsonian Renormalization Group Equation}},\
  }\href {https://doi.org/10.1016/j.aop.2008.08.008} {\bibfield  {journal}
  {\bibinfo  {journal} {Annals Phys.}\ }\textbf {\bibinfo {volume} {324}},\
  \bibinfo {pages} {414} (\bibinfo {year} {2009})},\ \Eprint
  {https://arxiv.org/abs/0805.2909} {arXiv:0805.2909 [hep-th]} \BibitemShut
  {NoStop}%
\end{thebibliography}%

\end{document}